\documentclass{aamod}

\usepackage{natbib}
\bibpunct{(}{)}{;}{a}{}{,}
\usepackage{graphicx}
\usepackage{txfonts}
\usepackage{lscape}

\begin{document}

\title{Solar low-lying cool loops and their contribution to the transition 
region EUV output}
\titlerunning{Low-lying cool loops and their contribution to the TR EUV output}
\author{C. Sasso\inst{1} \and V. Andretta\inst{1} \and D. Spadaro\inst{2} \and
R. Susino\inst{2}}

\offprints{C. Sasso, \email{csasso@oacn.inaf.it}}

\institute{INAF-Osservatorio
Astronomico di Capodimonte, Salita Moiariello 16, I-80131 Napoli,
Italy \and INAF-Osservatorio
Astrofisico di Catania, Via S. Sofia 78, I-95123 Catania,
Italy}

\date{Received / Accepted}

\abstract{}{In the last 30 years, the existence of small and cool magnetic 
loops (height $\lesssim8$~Mm, $T\lesssim10^5$~K) has been  proposed and
debated to explain the increase of the DEM (differential emission measure)
towards the chromosphere.}{We present hydrodynamic simulations of low-lying 
cool loops to study their conditions of existence and stability, and their 
contribution to the transition region EUV output.}{We find that stable, 
quasi-static cool loops (with velocities $<1$~km/s) can be obtained under 
different and more realistic assumptions on the radiative losses function with 
respect to previous works. A mixture of the DEMs of these cool loops plus 
intermediate loops with temperatures between $10^5$ and $10^6$~K can reproduce 
the observed emission of the lower transition region at the critical turn-up 
temperature point ($T\sim2\times10^5$~K) and below $T=10^5$~K.}{} 


\maketitle

\section{Introduction}

In the last 30 years, different theories have been proposed and debated to
explain the origin of the solar EUV output at temperatures below 1~MK. 
The idea that the transition region (hereafter TR) emission originates 
from the bases of the hot large-scale coronal loops is not confirmed by the 
measured DEM (Differential Emission Measure) which is orders of 
magnitude larger than predicted \citep{gabriel,athay}. The excess 
observed emission compared to the predictions of the traditional 
conduction-dominated TR picture has led, in particular, to the suggestion
that much of the TR plasma is confined in relatively small and cool magnetic
loops (height $\lesssim8$~Mm, $T\lesssim10^5$~K), strongly connected to the
chromosphere, but thermally insulated from the corona 
\citep{dowdy1,dowdy2,feldman1,feldman4}. 

The existence of this class of magnetic loops, predicted for decades, remains
far from being established. From the observational point of view, they are
indeed very difficult to observe. As most of their emission is in the UV,
traditional chromospheric diagnostics from ground, such as \ion{H}{$\alpha$}, 
only carry information about their lower boundary. On the other hand, the lack 
of spectral information with proper temperature sensitivity and spatial 
resolution from the current space-born EUV instruments capable of observing 
lines forming between $10^4$ and $10^5$~K has so far prevented any firm 
conclusions. 

Some recent observations with the VAULT instrument \citep[Very High Angular
Ultraviolet Telescope,][]{vault} in the \ion{H}{i} Ly-$\alpha$ line at very 
high spatial resolution, showed loop-like structures with estimated 
temperatures and densities ($T=10^4-3\times10^4$~K, 
$P=0.1-0.3$~dyne~cm$^{-2}$) that could be appropriate for the 
low-temperature end of cool loops \citep{patsourakos,vourlidas}. While such 
interpretation has been debated \citep{judge}, those observations, together 
with the persistence of the problem of the excess emission in EUV lines below 
$10^5$~K, call for further investigations on the matter. 
 
The general properties of static cool loops were first discussed by 
\citet{hood} and then studied more in detail from a theoretical point of view
by \citet{an86}, who demonstrated analytically that a mixture of static cool
loops with different temperatures can account for the observed lower
transition region DEM. This class of loops has specific characteristics
compared to the better studied coronal loops. While large-scale hot coronal
loops have dimensions of the order of $\sim10$~Mm, temperatures higher than 
$10^6$~K, and obey the ``static'' scaling laws described by 
\citet[hereafter RTV]{rtv}, cool loops are low-lying (estimated heights of the 
order of 1.1--5~Mm), nearly isobaric, and with maximum temperature below 
$10^5$~K. Another major difference with respect to classical coronal loops is 
in the role played by the three terms appearing in the hydrodynamical equation 
of plasma energy conservation: the heating rate, the conductive flux, and the 
radiative losses. Static cool loops are in approximate balance between the 
heating rate and radiative losses. Thus, inversely to coronal loops, the 
conductive flux plays a negligible role. 

\begin{figure*}
\centering
\includegraphics[clip=true,width=13cm]{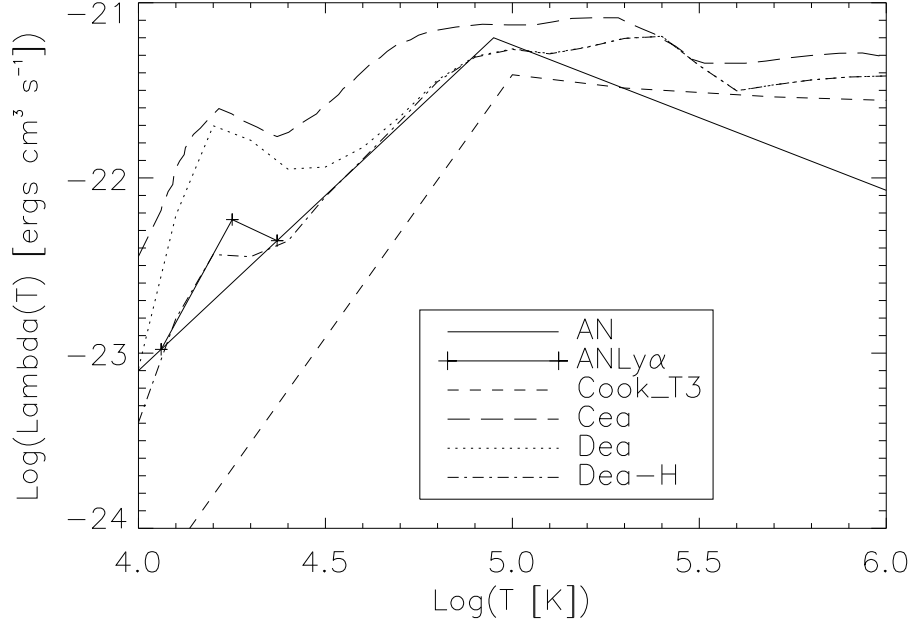}
\caption{Radiative losses functions considered in this work. Solid line: 
power-law segments function, equal to $~T^2$ (for log$T<4.95$~K) and $~T^{-1}$ 
(for log$T>4.95$~K); solid line plus star symbols: a peak mimicking the H 
Ly-$\alpha$ losses has been added to the previous function; dashed line: from 
\citet{cook}, but with a $T^3$ dependence below $T=10^5$~K; long-dashed line: 
from \citet{colgan}; dotted line: from \citet{chianti}; dot-dashed line: from 
\citet{chianti} without the H contribution.}
\label{fig:radlos}
\end{figure*}

\cite{cally}, following the study of \citet{an86}, treated in more detail the
stability, structure and evolution of these static cool loops, both 
analytically and numerically. They reveal, in particular, that the shape of 
the radiative losses function, $\Lambda(T)$, poses restrictive conditions on 
the existence of static cool loops. In the case of a realistic radiative 
losses function with the presence of the \ion{H}{i} Ly-$\alpha$ losses peak 
around $T=2\times10^4$~K (e.g., \citealt{chianti} or \citealt{colgan}, dotted 
and long-dashed line in Fig.~\ref{fig:radlos}, respectively), strictly static 
cool loops do not exist (see the next sections for details). It is important to
note that their analysis is limited to a minimum temperature of log$T=4.29$~K,
so that they cannot consider the existence of loops with maximum temperature 
below this value and cannot deal with Ly-$\alpha$ losses properly (even in the
thin losses case). They obtain static cool loops numerically with maximum 
temperature around $T=8\times10^4$~K, assuming a $T^3$ dependence of the 
radiative losses function on the temperature below $T=10^5$~K (dashed line in 
Fig.~\ref{fig:radlos}). They conclude, however, that this temperature is a 
factor 2--3 too low to explain the observed low-TR emission, even if they do 
not show any calculated DEM function to support these conclusions. Moreover,
they state that ``the steep $T^3$ radiative losses function even if so
beneficial for the existence and stability of cool loops rests on shaky 
foundations''.  

After the cited work of \citet{cally}, our paper is the first attempt to study 
numerically the conditions of existence of small and cool loops (height 
$\lesssim5$~Mm, $T\lesssim10^5$~K) and, when they exist, their physical 
structure and their contribution to the TR emission. We show that stable, 
quasi-static cool loops (with plasma velocities $<1$~km/s) can be obtained 
through hydrodynamic simulations also assuming different forms of 
$\Lambda(T)$, with respect to the work of \cite{cally}. 

\citet{peter,dem} made the first successful attempt to reproduce the 
shape of the DEM curve quantitatively and qualitatively, even at temperatures 
below $\log T=5.3$~K. They used a forward model in which they synthesize 
spectra from three-dimensional MHD simulations of the whole Sun atmosphere, 
from the chromosphere to the corona. We think that their results do not 
exclude ours. Their simulations may include structures that could be related
to the kind of loops we are studying. However, the loops we describe in this 
paper would be covered by only very few resolution elements in their 
simulation, and in any case resolving the gradients and the dynamics of the 
relevant quantities in our loop models would require a much higher
resolution. Therefore, we regard our study as complementary to the kind of 
large-scale simulations by \citet{dem}.

In the course of our study of loops with the properties just described, we 
have also found low-lying quasi-static loops with temperatures in the range
$10^5-10^6$~K. Following one of the latest loop classifications 
\citep{realesolo} we should refer to these loops also as ``cool loops''. In
order to avoid confusion we will refer to them as ``intermediate temperature
loops''. This class of loops was first studied by \citet{foukal} and observed 
by \citet{brekke}. From observations, they are generally detected in the UV 
lines and appear to be steady for long times even if more variable and dynamic 
then hot coronal loops, probably due to the presence of substantial flows 
\citep{brekke,digiorgio,realesolo}. Their estimated densities are in the range 
$10^9-10^{10}$~cm$^{-3}$ \citep[e.g.,][]{brown}. Even if observed intermediate 
temperature loops are often related to flows we decided to report however in 
this paper our findings on their existence (at least analytically) in a 
quasi-static state. Studying the role of flows in the conditions of their
existence would mean introducing a new dimension in the parameter space to be 
explored, and this is beyond the scope of this paper, concentrated on 
quasi-static cool loops. Anyway, the quasi-static intermediate temperature 
loops we find have different physical properties with respect to observed 
loops with that temperature and this may be correlated with flows.    

The paper is structured as follows: in Sec.~\ref{sec:model}, we describe the 
numerical model, and the different radiative losses functions adopted in this 
work are introduced. In Sec.~\ref{sec:results}, the hydrodynamic simulations 
and the different loops obtained (cool and intermediate temperature loops) are 
presented and their properties discussed and analyzed. Section~\ref{sec:dem}, 
in particular, is dedicated to the calculated DEMs of these loops and to the 
comparison with the observed one. Finally, in the conclusions 
(Sec.~\ref{sec:concl}), the role of the cool and intermediate temperature 
loops in the solar atmosphere and the possibility to observe them with current 
instrumentation will is treated.

\section{Numerical modelling and calculations}\label{sec:model}

The set of hydrodynamic equations for mass, momentum and plasma energy 
conservation for a fully ionized hydrogen plasma have been solved in a 
unidimensional, magnetically confined loop of constant cross-section with
ARGOS, a 1-D hydrodynamic code with the fully adaptive-grid package PARAMESH 
\citep{ARGOS,PARAMESH}. A fully adaptive-grid is necessary to adequately
resolve one or more evolving regions of steep gradients, for example the thin 
segments of the transition region chromosphere-corona of the loop (for coronal
loops) or shocks in dynamical simulations. The geometry of the loop is 
determined by the following analytic form: 
\begin{eqnarray}\label{eqn:loop1}
z(s)=h-\frac{h\left(1-\sqrt{1-(s/\gamma)^2}\right)}{1-\sqrt{1-(L/2\gamma)^2}}, 
\end{eqnarray}
where $z$ is the height above the chromospheric footpoints, $s$ is the 
curvilinear coordinate along the field lines, $h$ is the height of the loop 
apex (i.e., $z_{max}$), $L$ is the loop total length, and the constant 
$\gamma$ is defined in terms of $h$ and $L$ as
\begin{eqnarray}\label{eqn:loop2}
\gamma=\frac{L-2h}{2\sqrt{1-4h/L}}. 
\end{eqnarray}
This form implies that $z(s)=0$ at the footpoints, $z(s)$ has its maximum $h$
at the midpoint, $s=0$, and $dz/ds=\pm 1$ when $s=\pm L/2$. 
Equations~\ref{eqn:loop1} and \ref{eqn:loop2} generate an arched loop of given 
length $L$ and apex height above the chromosphere $h$. 

The numerical model includes a thick chromosphere at each footpoint $26.7$~Mm 
deep, acting as a mass reservoir; the total length of the flux tube is 
therefore $L+53.4$~Mm. The boundary conditions imposed are at the two 
endpoints (rigid wall, fixed temperature), which are located at 
$s=\pm(L/2+26.7)$~Mm. The temperature of this chromosphere is set at 
$T=10^4$~K. This is something new with respect to previous works, in which the 
chromospheric temperature was set to $T=2$ or $3\times10^4$~K 
\citep[e.g.][]{an86,cally,spadaro1}. Having a chromospheric temperature at 
$10^4$~K allows the inclusion of the full peak of the \ion{H}{i} Ly-$\alpha$ 
line in the optically thin radiative losses function and to explore the 
temperature range ($10^4-3\times10^4$~K) observed by VAULT. The inclusion of
the Ly-$\alpha$ as an optically thin line is of course only a first step, 
suitable for this exploratory study. Work is in progress to take into account 
the transfer of Ly-$\alpha$ radiation, as well as to include partial plasma 
ionization at the lower temperature end of the simulations. Since we take, by 
definition, the beginning of the chromosphere as the level at which the plasma 
drops below $10^4$~K, the exact position of the top of the chromosphere 
($s=\pm L_\mathrm{i}/2$ at the beginning of the simulation) changes during the 
calculation with the plasma filling or evacuating the loop. So, at end of the 
simulation, we will have a new position for the top of the chromosphere 
$s=\pm L_\mathrm{f}/2$ and, consequently, a new value of $h=h_\mathrm{f}$, 
where $h_\mathrm{f}$ is no longer the  geometrical parameter defining the 
shape of the loop, but the height of the  loop apex above the $T=10^4$~K level. 

The set of conservation equations for mass, momentum, and energy, 
respectively, in a one-dimensional plasma solved by ARGOS is
\begin{eqnarray}
\frac{\partial}{\partial t}\rho+\frac{\partial}{\partial s}(\rho v)&=&0, \label{eq:1}\\
\frac{\partial}{\partial t}(\rho v)+\frac{\partial}{\partial s}(P+\rho v^2)&=&-\rho g_{\parallel}(s), \label{eq:2}\\
\frac{\partial U}{\partial t}+\frac{\partial}{\partial s}\left(Uv+F_{\textmd{\tiny{c}}}\right)&=&-P\frac{\partial}{\partial s}v+E(s,t)-n^2\Lambda(T), \label{eq:3}\\
F_{\textmd{\tiny{c}}}&=&-10^{-6}T^{5/2}\frac{\partial}{\partial s}T.\label{eq:condflux}
\end{eqnarray}
where $t$ is the time, $\rho$ the mass density, $v$ the velocity, $P$, $T$
and $n$ are the gas pressure, temperature, and electron number density, 
respectively, linked by the equation of state $P=2nkT$, with $k$ the Boltzmann 
constant; $U=3P/2$ is the internal energy, $s$ the curvilinear coordinate 
along the loop, $E(s,t)$ the assumed form for the coronal heating rate, 
$n^2\Lambda(T)$ the plasma optically thin radiative losses, with $\Lambda(T)$ 
the radiative losses function, $g_{\parallel}(s)$ the component of the solar 
gravity along the loop axis, and $F_{\textmd{\tiny{c}}}$ the thermal
conductive flux, in CGS units. 

\subsection{Radiative losses functions}

The hydrodynamic simulations are performed assuming different shapes for the 
radiative losses function, that are shown in Fig.~\ref{fig:radlos}. Since 
radiative losses were hard-coded in the original version of ARGOS, we modified 
the code to allow for arbitrary, tabulated $\Lambda(T)$. We have verified with 
some benchmark tests, that the modification does not introduce significant 
overheads in the calculations. First, we use the same approximation of the 
radiative losses function adopted by 
\citet[][solid line in Fig.~\ref{fig:radlos}, $\Lambda_{AN}$ in 
Table~\ref{tab:1}]{an86}, who fitted the data of \citet{gaetz}, assuming that 
the $\Lambda(T)$ can be approximated by an increasing power-law of temperature 
for $\log T<4.95$~K ($\propto T^a$), and by a decreasing power-law for 
$\log T>4.95$~K ($\propto T^{-b}$), with $a$ and $b$ both positive. In 
particular:
\begin{eqnarray}
\Lambda_{AN}(\mbox{T})= 
\left\{
\begin{array}{rl}
7.94\times10^{-32} T^2 & \mbox{for log}T<4.95 \mbox{ K} \\
5.62\times10^{-17} T^{-1} & \mbox{for log}T>4.95 \mbox{ K}
\end{array}
\right.
\end{eqnarray}
This approximation does not consider the peak due to the H Ly-$\alpha$ 
radiative losses at around $T=2\times10^4$~K. In order to study its effect on 
the simulations, we added to the function $\Lambda_{AN}$ a peak mimicking the H 
Ly-$\alpha$ losses (solid line plus cross symbols in Fig.~\ref{fig:radlos}, 
$\Lambda_{ANLy\alpha}$ in Table~\ref{tab:1}). The peak has approximately the 
same shape of the one in the realistic radiative losses function 
$\Lambda_{Cea}$, that we are going to describe. It was built by taking the two 
lines parallel to the ascending and to the descending branches of the 
Ly-$\alpha$ peak present in the function $\Lambda_{Cea}$, and making them pass 
through the point with abscissa $\log T=4.21$~K, corresponding to the peak top 
value.

The long-dashed line and the dotted line in Fig.~\ref{fig:radlos} are two 
realistic radiative losses functions. The first is from the work of 
\citet[][$\Lambda_{Cea}$ in Table~\ref{tab:1}]{colgan} and the other was 
obtained using the ``CHIANTI'' database, version 6 \citep[][$\Lambda_{Dea}$ in 
Table~\ref{tab:1}]{chianti}, adopting the ionization equilibrium of 
\citet{ioneq}, and a coronal mixture of elements. We choose these two
functions because they are the most recent and are based on the latest atomic 
data available. A discussion on the differences, quite significant, between 
the two functions can be found in \citet{chianti}. 

The last radiative losses function used in this paper has been obtained by 
removing the contribution of hydrogen from the function $\Lambda_{Dea}$ 
(dot-dashed line Fig.~\ref{fig:radlos}, $\Lambda_{Dea-H}$ in 
Table~\ref{tab:1}). Even if the hydrogen losses have been subtracted, a 
smaller peak remains at around $T=2\times10^4$~K due to other elements losses 
(mainly helium). 

In the next section, the results from simulations using these different
radiative losses functions will be introduced. The list of simulated loops,
together with the relevant parameters, is given in Table~\ref{tab:1}.   

\section{Results and discussion}\label{sec:results}

\begin{figure*}
\centering
\includegraphics[clip=true,width=6.5cm]{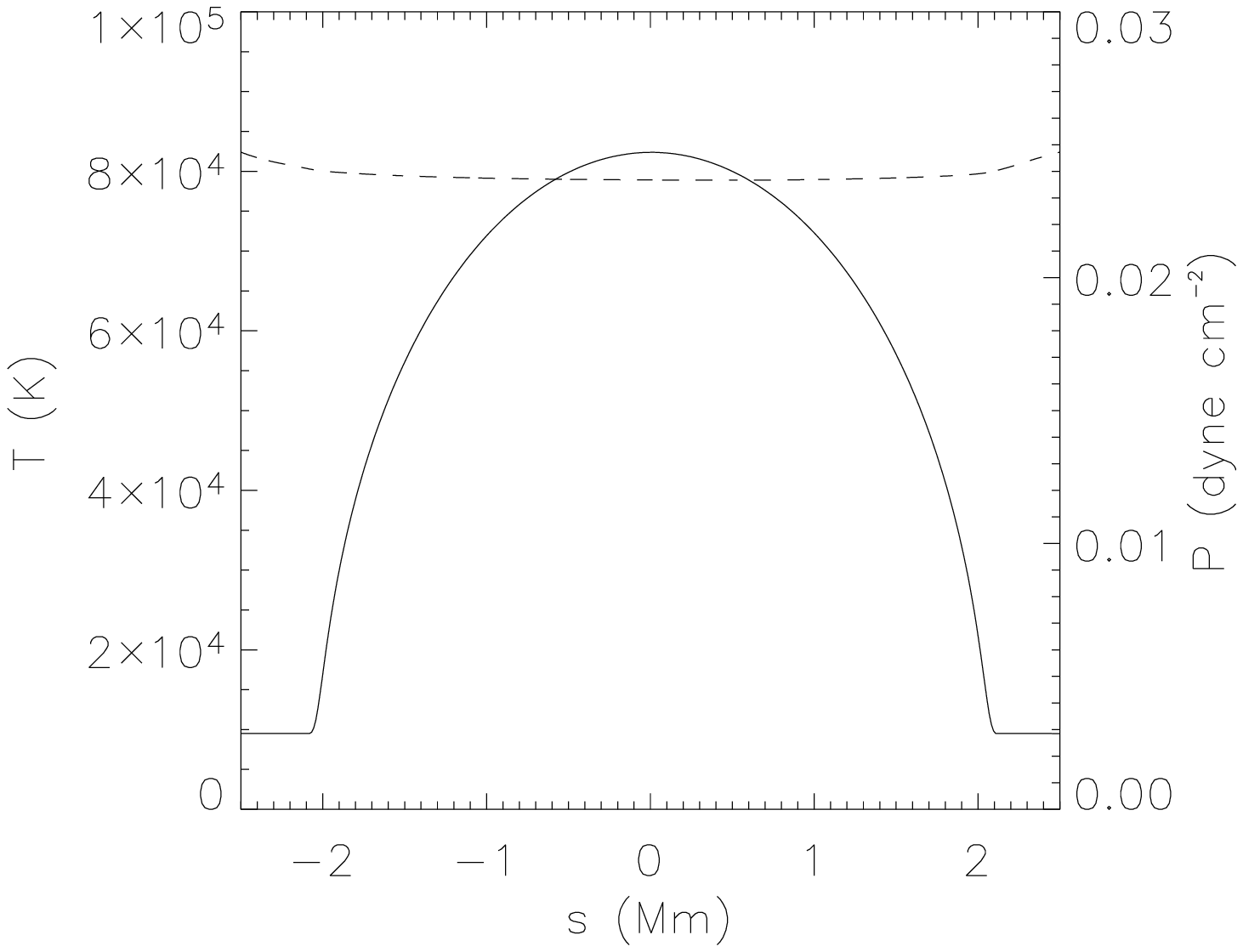}
\includegraphics[clip=true,width=6.5cm]{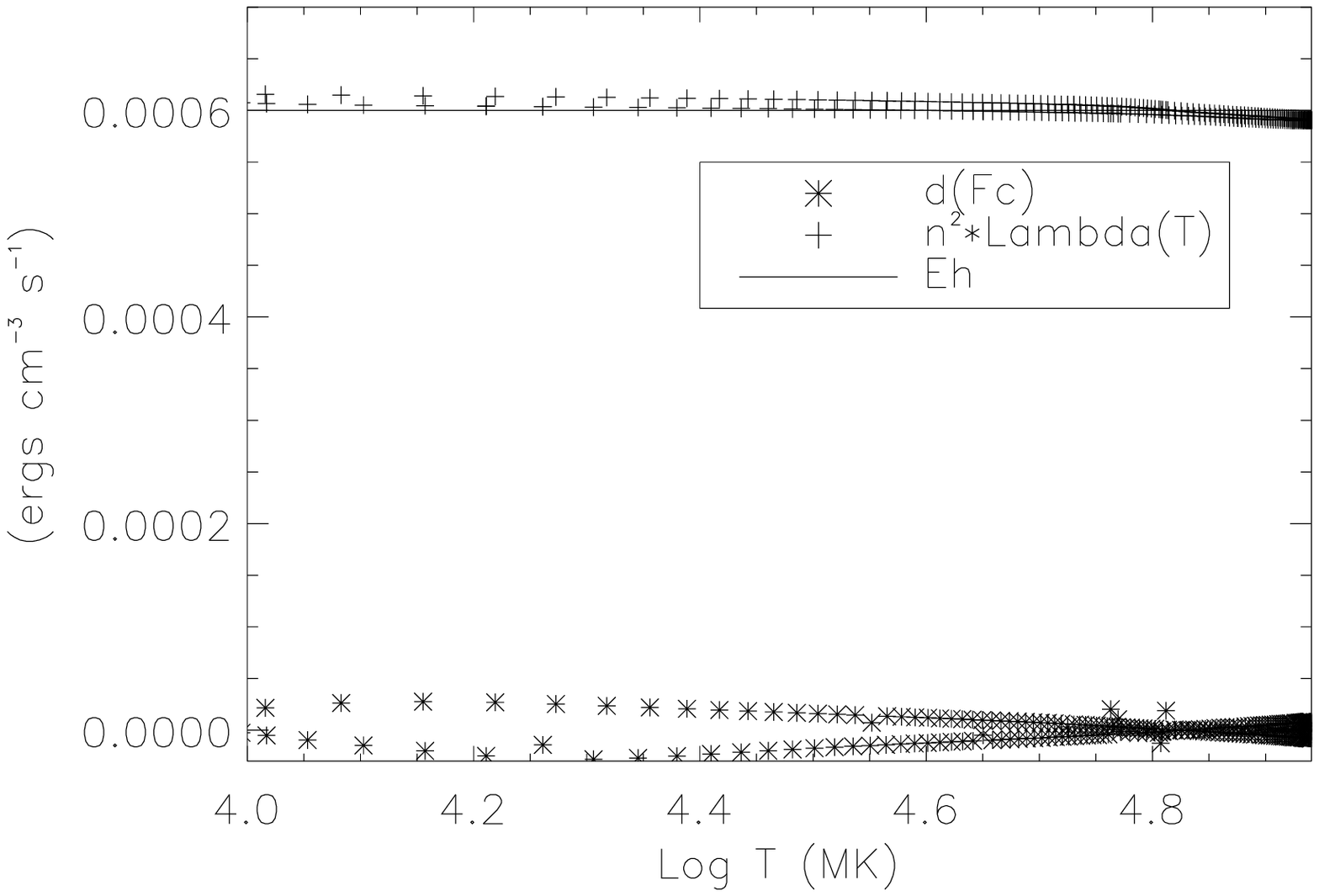}
\includegraphics[clip=true,width=6.5cm]{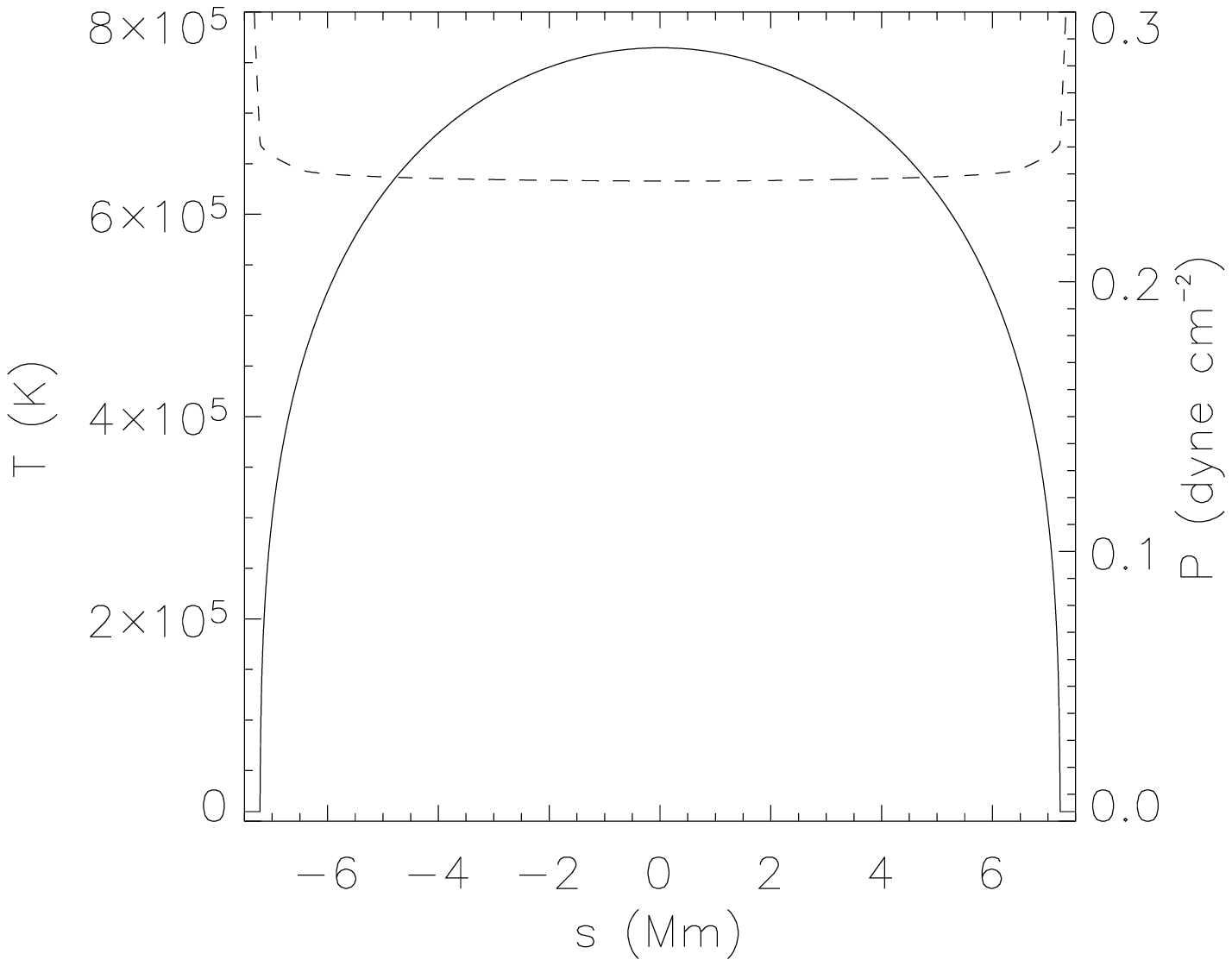}
\includegraphics[clip=true,width=6.5cm]{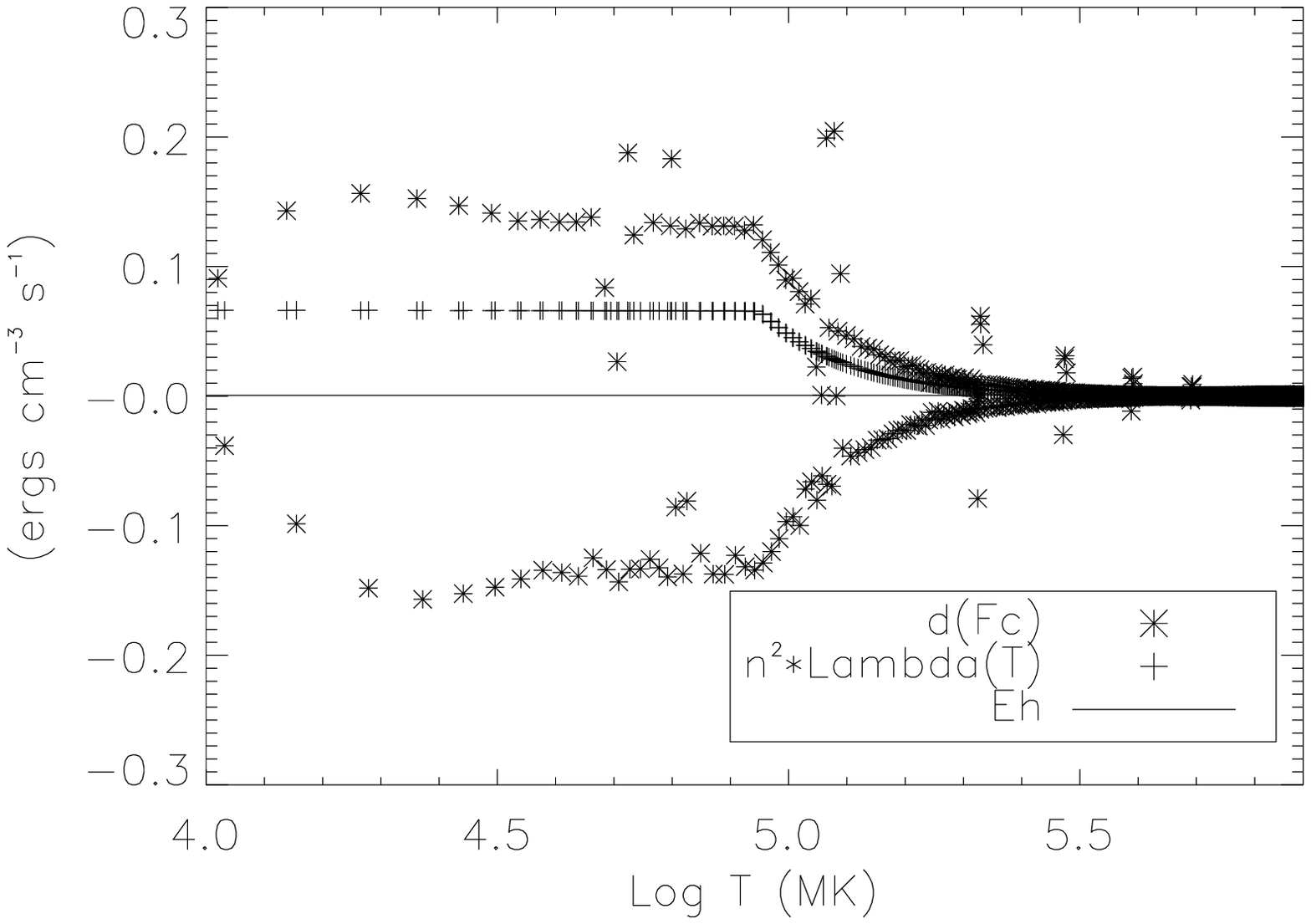}
\includegraphics[clip=true,width=6.5cm]{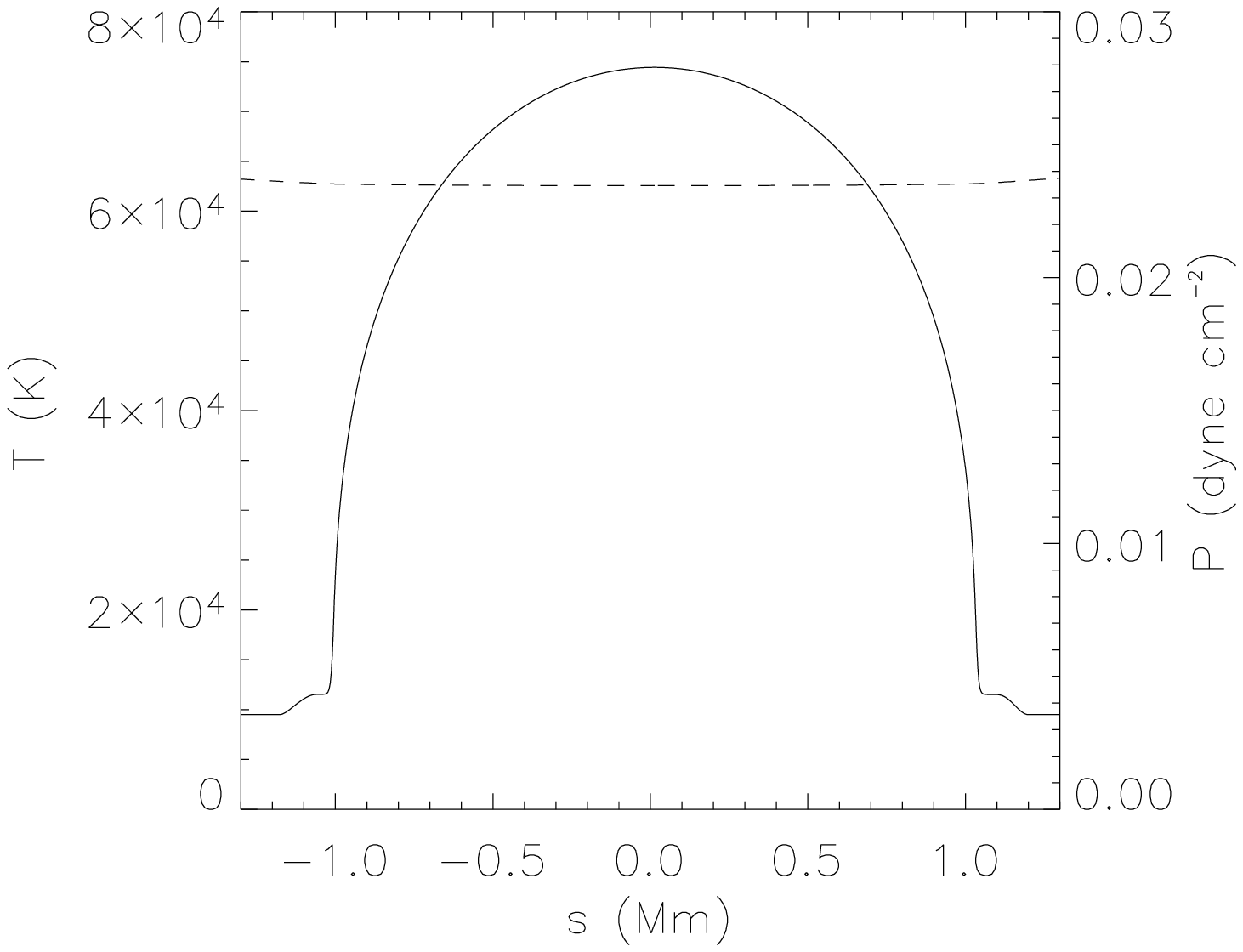}
\includegraphics[clip=true,width=6.5cm]{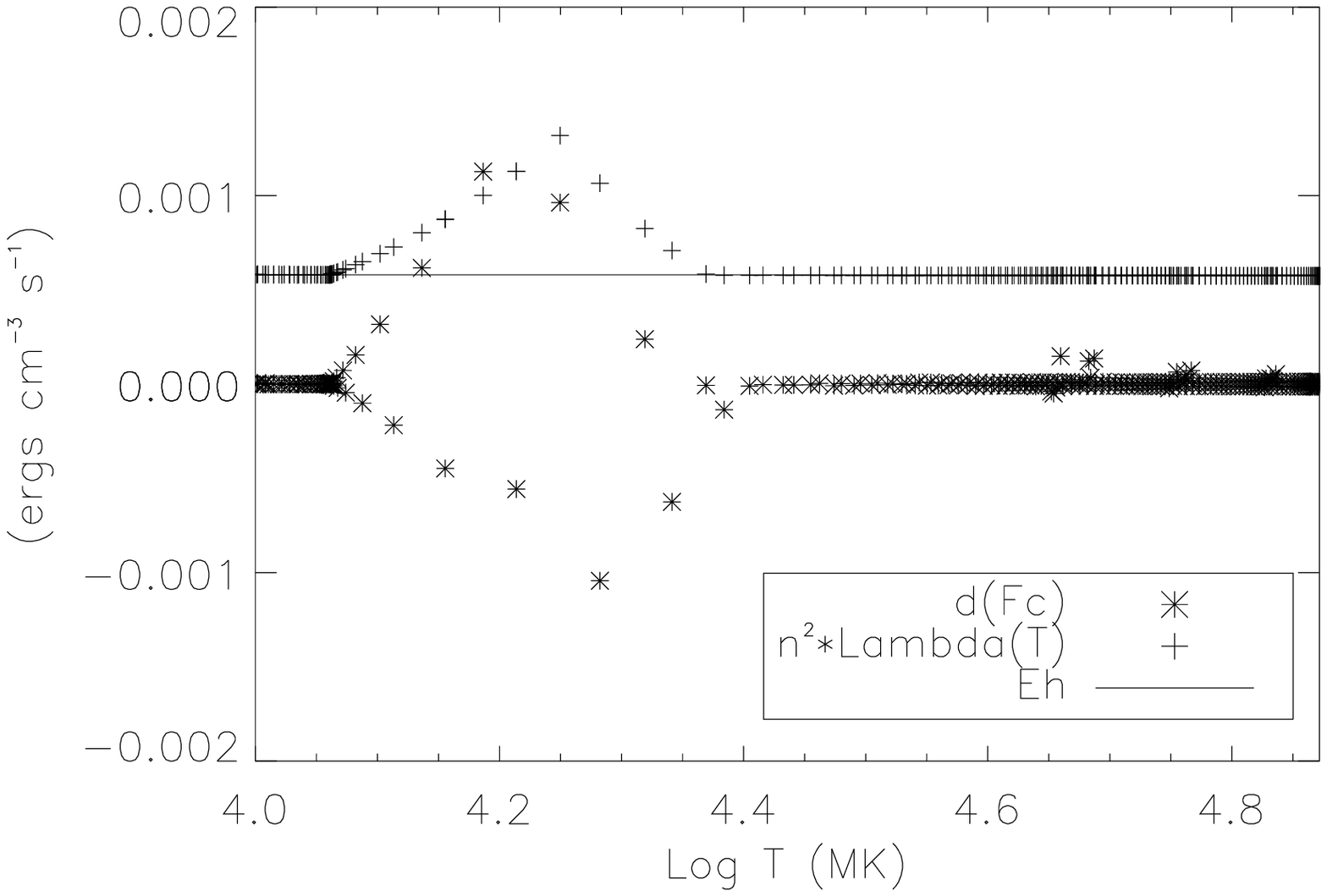}
\caption{Top panel, left: temperature (solid line) and pressure (dashed line) 
as a function of the curvilinear coordinate along the field lines, $s$. Right: 
divergence of the conductive flux (asterisks), radiative losses (crosses) and 
constant heating rate, $E_{\textmd{\tiny{h}}}$ (solid line), as a function of 
the temperature, for loop 4. Middle and bottom panels: as in the top panels 
for loop 10 and 15, respectively.}
\label{fig:TP161109}
\end{figure*}
\begin{figure*}
\centering
\includegraphics[clip=true,width=6.5cm]{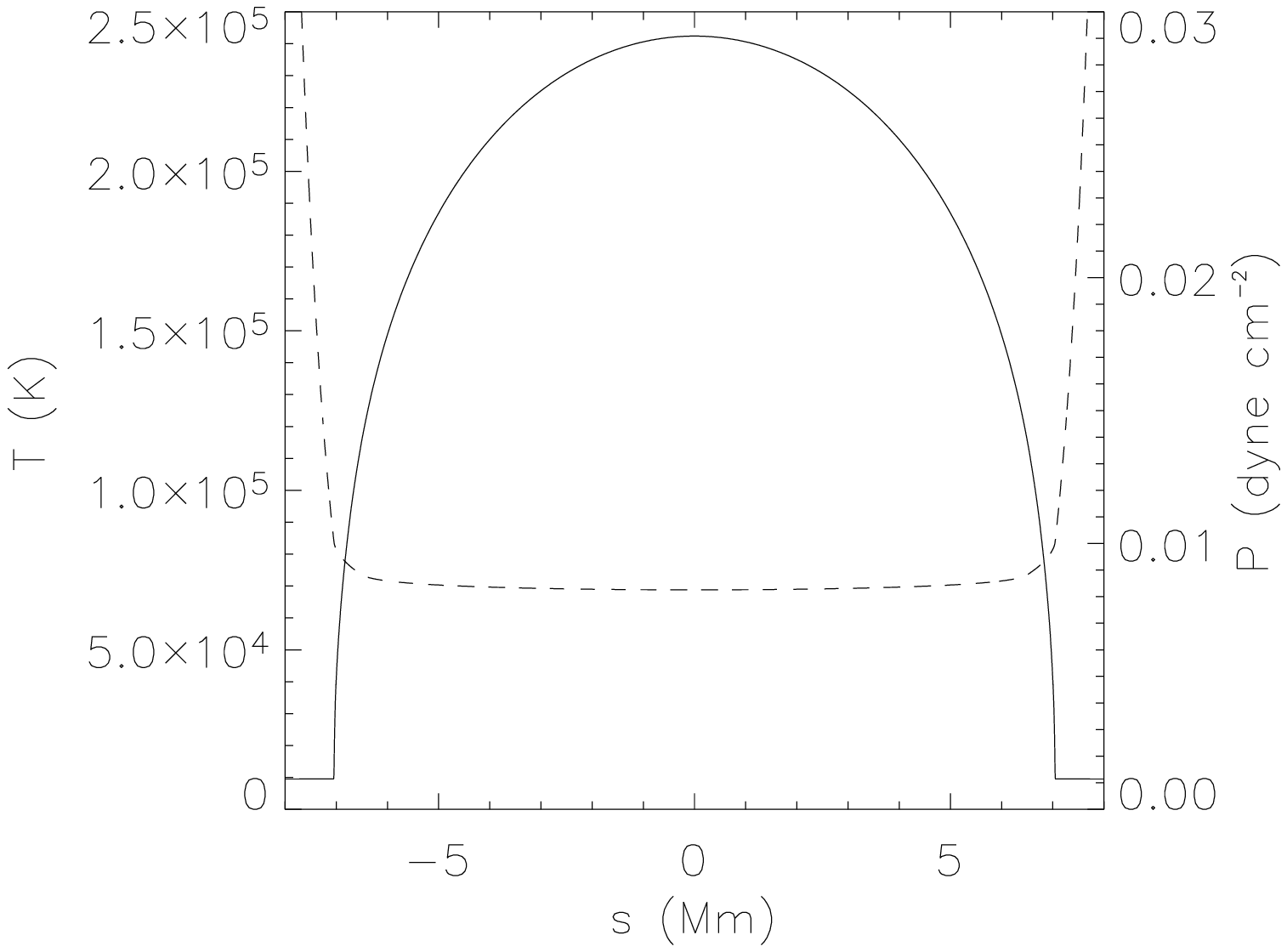}
\includegraphics[clip=true,width=6.5cm]{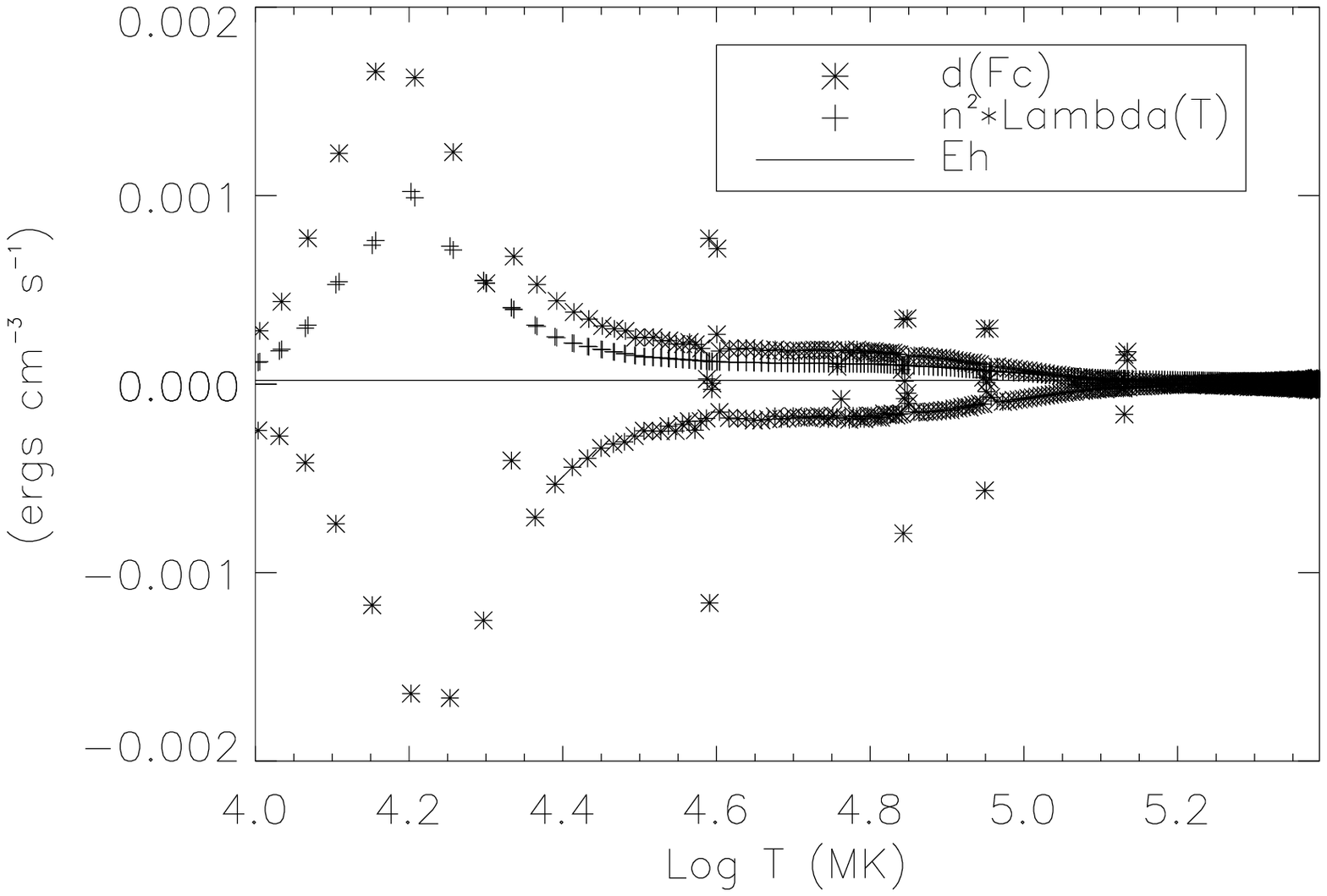}
\includegraphics[clip=true,width=6.5cm]{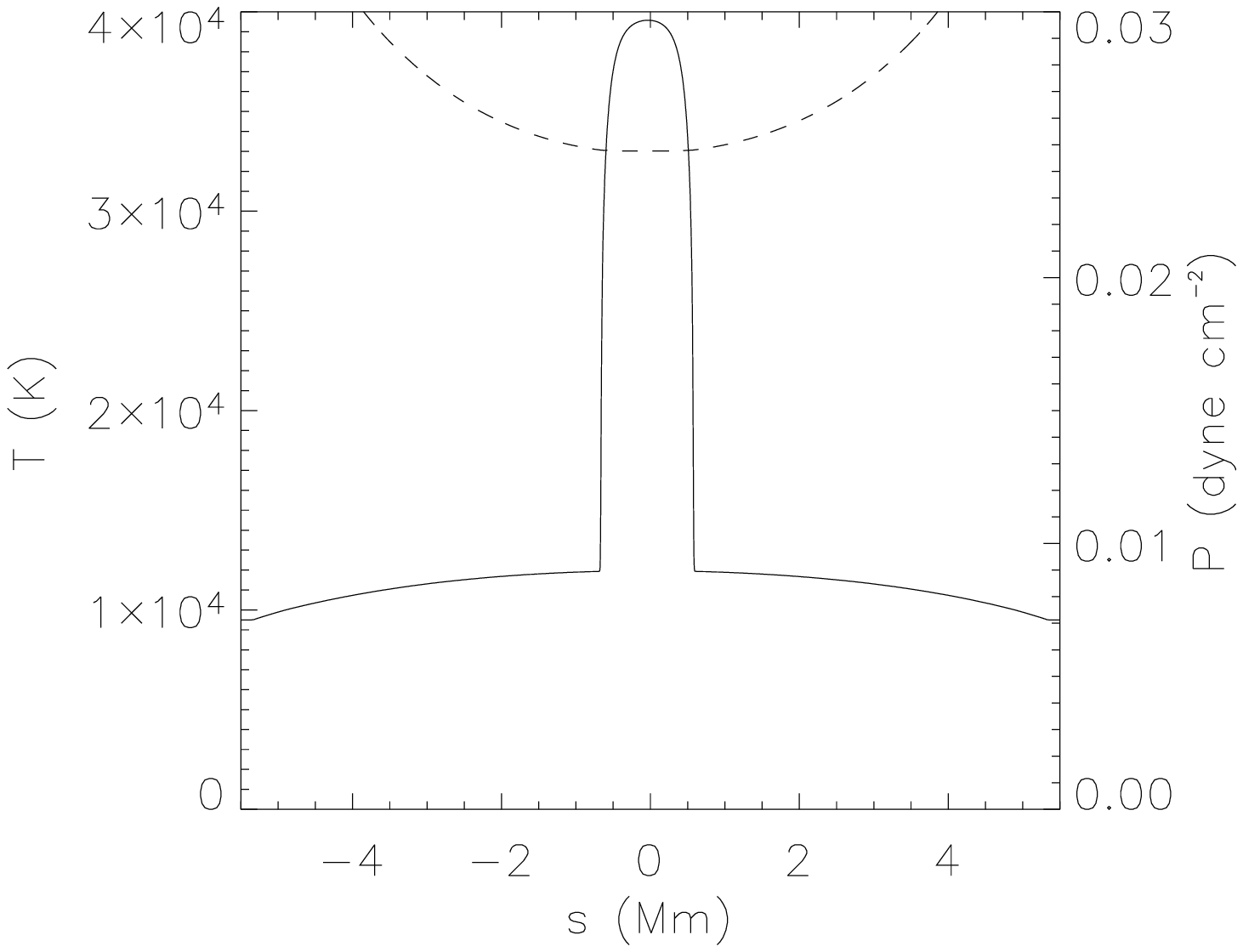}
\includegraphics[clip=true,width=6.5cm]{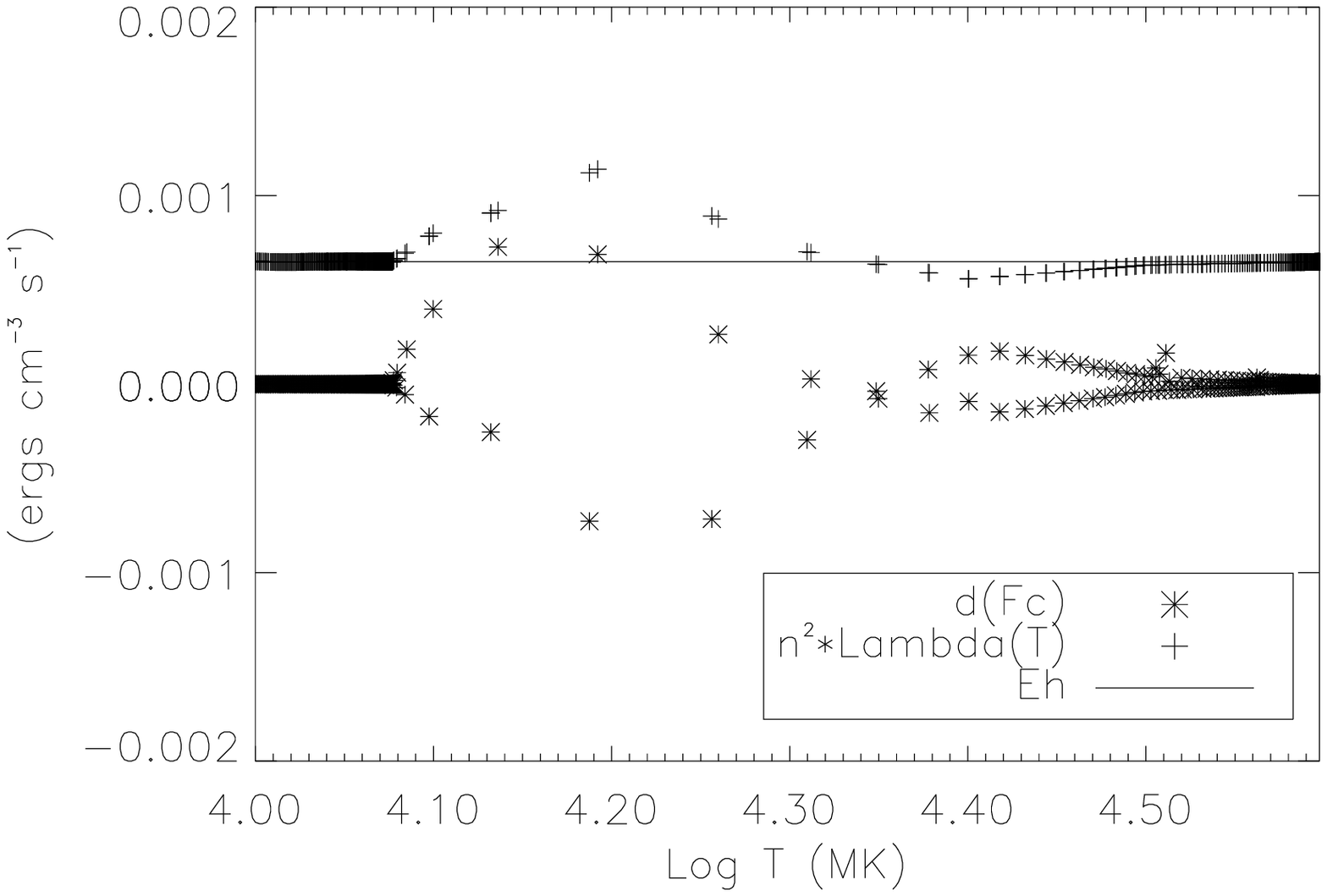}
\caption{As in Fig.~\ref{fig:TP161109} for loop 17 (top panels) and loop 25
  (bottom panels).}
\label{fig:TP081010}
\end{figure*}
\begin{table*}
\caption{Loops parameters.}
\label{tab:1}
\centering
\small
\begin{tabular}{ccccccc}
  \hline\hline
  Loop&$\Lambda(T)$&$E_{\textmd{\tiny{h}}}$&$T_{max}$&$P$&$L/2$&$h$\\
  &&$10^{-4}$~ergs~cm$^{-3}$~s$^{-1}$&MK&dyne cm$^{-2}$&Mm&Mm\\
  \hline
  \multicolumn{7}{c}{Cool loops - $\Lambda_{AN}$} \\ 
  1  &AN        &4    &0.026   &0.019  &0.7 &0.005 \\
  2  &AN        &4.5  &0.078   &0.021  &2   &0.04  \\
  3  &AN        &5.5  &0.063   &0.023  &1.6 &0.02  \\
  4  &AN        &5.9  &0.083   &0.024  &2.1 &0.04  \\  
  5  &AN        &7.1  &0.057   &0.026  &2.5 &0.04  \\  
  6  &AN        &7.5  &0.074   &0.027  &1.8 &0.02  \\
  \hline
  \multicolumn{7}{c}{Intermediate temperature loops - $\Lambda_{AN}$} \\   
  7  &AN        &5.5  &0.740   &0.22   &7.2 &1.32  \\
  8  &AN        &5.5  &0.812   &0.24   &8.4 &2.52  \\  
  9  &AN        &5.5  &0.846   &0.25   &8.9 &3.02  \\
  10 &AN        &6    &0.765   &0.24   &7.2 &1.32  \\
  11 &AN        &6    &0.774   &0.24   &7.1 &1.22  \\
  12 &AN        &6    &0.870   &0.27   &9   &3.12  \\
  13 &AN        &8    &0.839   &0.31   &7.4 &1.52  \\
  \hline
  \multicolumn{7}{c}{Cool loops - $\Lambda_{ANLy\alpha}$} \\   
  14 &ANLy$\alpha$ &5.5  &0.012   &0.023  &0.6 &0.003 \\
  15 &ANLy$\alpha$ &5.8  &0.074   &0.023  &1.2 &0.01  \\
  16 &ANLy$\alpha$ &5.9  &0.082   &0.024  &1.4 &0.02  \\
  \hline
  \multicolumn{7}{c}{Intermediate temperature loops - $\Lambda_{Dea/Cea}$} \\   
  17 &Dea       &0.2  &0.242   &0.008  &7   &1.12  \\
  18 &Cea       &0.2  &0.261   &0.008  &7   &1.12  \\
  19 &Cea       &1.01 &0.461   &0.03   &7.8 &1.92  \\
  20 &Cea       &6    &0.684   &0.13   &6.9 &1.02  \\
  21 &Cea       &6.47 &0.702   &0.13   &6.9 &1.02  \\
  22 &Cea       &8    &0.745   &0.16   &7   &1.12  \\
  23 &Cea       &8    &0.748   &0.16   &7   &1.12  \\  
  \hline
  \multicolumn{7}{c}{Cool loops - $\Lambda_{Dea-H}$} \\   
  24 &Dea-H     &6    &0.012   &0.024  &5.2 &0.27  \\
  25 &Dea-H     &6.5  &0.040   &0.025  &5.3 &0.29  \\
  26 &Dea-H     &7.4  &0.050   &0.026  &5.5 &0.32  \\
  \hline
  27 &AN$10^5$  &6   &0.087   &0.024  &2.3 &0.04\\
\hline
\end{tabular}
\end{table*}
\begin{figure*}
\centering
\includegraphics[clip=true,width=13cm]{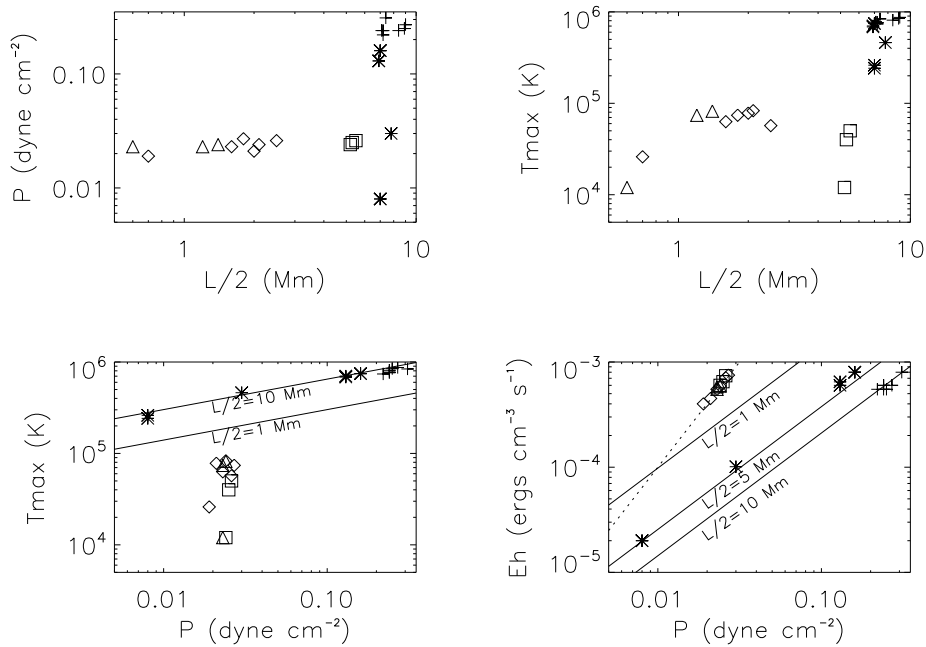}

\caption{Behavior of the physical parameters for loops 1--6 (diamonds), 7--13 
(crosses), 14--16 (triangles), 17--23 (asterisks) and 24--26 (squares) of 
Tab.~\ref{tab:1}. The solid lines represent the RTV scaling laws for coronal
loops, for different values of $L/2$. The dotted line represents 
$E_{\textmd{\tiny{h}}}=P^2$.}
\label{fig:leggiscala}
\end{figure*}
\begin{figure*}
\centering
\includegraphics[clip=true,width=13cm]{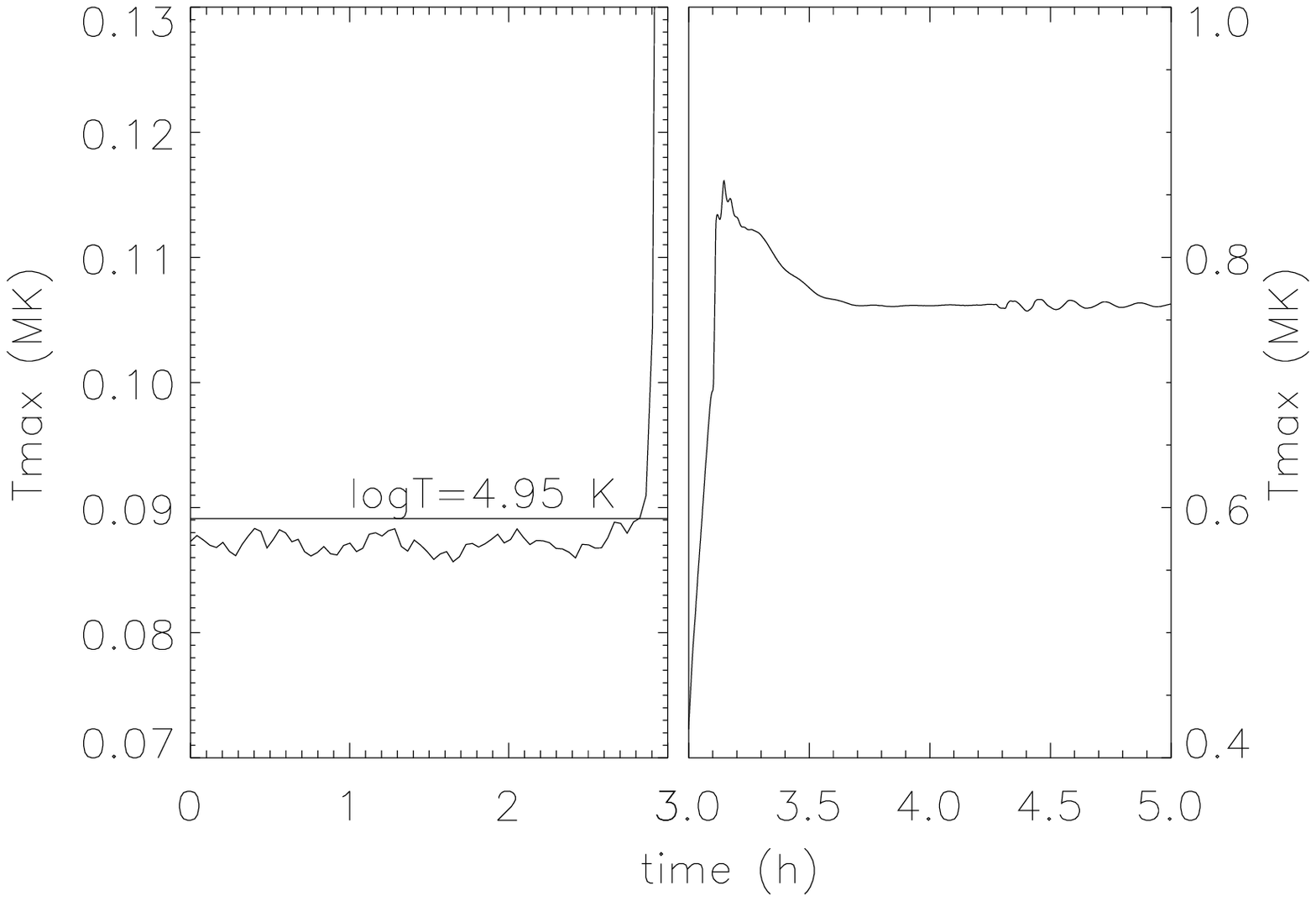}
\caption{Evolution of the maximum temperature of loop 11 in Tab.~\ref{tab:1}
during the hydrodynamic simulation. The horizontal solid line indicates the 
break temperature log$T=4.95$~K.}
\label{fig:restart}
\end{figure*}
To derive the initial equilibrium, we followed the approach of 
\citet{spadaro1}, i.e., we simply started with a discontinuous temperature 
profile in which $T=T_\mathrm{i}$ (with $T_\mathrm{i}$ a constant value) in
the corona and $T=10^4$~K in the chromosphere. The density was constant in the 
corona, $n=n_\mathrm{i}$, and increased exponentially with depth in the 
chromosphere at the appropriate scale height. Hence, the plasma was in 
approximate force balance to start with, but not in thermal equilibrium.
During the simulation, the loop evolves towards a stable quasi-static solution
of the hydrodynamic equations (Eqs.~\ref{eq:1}--~\ref{eq:3}), a solution, that
is, that approximately fulfills the static and stationary versions of those 
equations:
\begin{eqnarray}
\frac{\partial}{\partial s}P&=&-m_{\textmd{\tiny{H}}}ng_{\parallel}(s),\label{eq:hydro} \\
\frac{\partial}{\partial s}F_{\textmd{\tiny{c}}}&=&E_{\textmd{\tiny{h}}}-n^2\Lambda(T),\label{eq:energy}
\end{eqnarray}
where we used $\rho=m_{\textmd{\tiny{H}}}n$ for a fully ionized hydrogen
plasma with $m_{\textmd{\tiny{H}}}$ the hydrogen mass. We consider a loop in a
quasi-static equilibrium state when the plasma velocities are lower than
1--2~km/s. With this definition, we find that in all our simulations the
thermodynamic parameters of a loop in quasi-static equilibrium oscillate
around the equilibrium value with an amplitude much smaller than 1\%. We took 
this quasi-static equilibrium state as a result or, in some cases (see below), 
as an initial equilibrium state ($t=0$~s) for a new simulation with different 
parameters (e.g.: different heating rate).

In all the simulations we assume spatially uniform and temporally constant 
heating rate per unit volume, $E(s,t)=E_{\textmd{\tiny{h}}}$, and we ensure
that they last long enough for the end time to be much larger than the loop 
radiative cooling time.

\subsection{Description of results}\label{subsec:description}

We ran numerous simulations, extensively exploring the parameter space, under 
different initial conditions. We list in Tab.~\ref{tab:1} and discuss below a 
representative selection of the loops in quasi-static equilibrium we have 
obtained. Loops 3-4, 7-13, and 17-23 are obtained by starting the simulation 
with a loop in approximate force balance but not in thermal equilibrium as 
described previously, using different combinations of $T_\mathrm{i}$ and 
$n_\mathrm{i}$ that we thought be appropriate for cool loops. All these 
simulations have an initial half length, $L_\mathrm{i}/2=6.5$~Mm and an 
initial height, $h_\mathrm{i}=0.62$~Mm. Once we obtained two quasi-static cool 
loops (loops 3 and 4), we used them as initial equilibrium states to perform 
new simulations. In this way, starting in particular from loop 3 and changing 
some parameters we have obtained all the other cool loops 1-2, 5-6, 14-16, and 
24-26. Loops 1 and 2 have been obtained by changing the heating rate and the 
initial height, $h_\mathrm{i}=0.7$~Mm. Loops 5 and 6 have been obtained by 
changing again the heating rate and the initial height but to the value 
$h_\mathrm{i}=0.54$~Mm. Finally, loops 14-16 and 24-26 have again different 
heating rates but also different radiative losses functions.  

\subsubsection{Loops from idealized $\Lambda(T)$}

The first group of simulations we have done assumes as a radiative losses 
function $\Lambda_{AN}$, in order to test the analytical results of 
\citet{an86}. We are able to obtain quasi-static cool loops with maximum
temperature between $\sim2.6$ and $8.3\times10^4$~K, using
$E_{\textmd{\tiny{h}}}$ in the range 
$4-7.5\times10^{-4}$~ergs~cm$^{-3}$~s$^{-1}$. In Table~\ref{tab:1} we report 
only some example of the cool loops found (loops 1--6). These loops have the 
properties analytically predicted by \citet{an86}: they are small 
($L/2=0.7-2.5$~Mm and $h=0.005-0.04$~Mm), nearly isobaric, and in approximate 
balance between the heating rate and radiative losses, 
$E_{\textmd{\tiny{h}}}=n^2\Lambda(T)\gg|\nabla
\mathbf{F}_{\textmd{\tiny{c}}}|$. Also the low pressure values, compared to 
coronal hot loops, were analytically predicted by \citet{an86}. According to 
their calculations, the pressure of a cool loop has to be at least one and 
one-half order of magnitude smaller then that of a hot loop with maximum 
temperature bigger than $10^6$~K. This result is confirmed by our calculations. 

In Figs.~\ref{fig:TP161109} and \ref{fig:TP081010} we plot the behavior of the 
loop parameters as well as of the terms of the energy equation~\ref{eq:energy} 
for five loops chosen as examples for each group of loops obtained with a 
different $\Lambda(T)$ (loops 4, 10, and 15 from top to bottom in 
Fig.~\ref{fig:TP161109}, and loops 17 and 25, from top to bottom in 
Fig.~\ref{fig:TP081010}). The left panels show the temperature (solid line) 
and the pressure (dashed line) profiles as a function of the curvilinear 
coordinate, $s$, while the right panels show the radiative losses energy term, 
$n^2\Lambda(T)$ (crosses), the heating rate, $E_{\textmd{\tiny{h}}}$ (solid 
line), and the divergence of the conductive flux, $\nabla 
\mathbf{F}_{\textmd{\tiny{c}}}$ (asterisks), as a function of the
temperature. The plots show the variation of loop temperature, density, and
energy terms along the loop at a particular instant (corresponding to the end 
time of the simulation). Since the loops are still in a quasi-static 
equilibrium state, with the plasma subject to motions (oscillations, for 
instance) of small, but not negligible amplitude, those profiles do not appear
exactly symmetric around the loop apex ($s=0$). Averaging profiles over an 
appropriate number of time steps would cancel out such small variations, thus 
yielding nearly symmetric profiles.
 
The maximum temperature of loop 4 (top left panel of Fig.~\ref{fig:TP161109}), 
chosen as the example for the first group of loops 1-6, is $8.3\times10^4$~K 
and the pressure is constant along the loop (within $1\%$ above the 
chromosphere). The terms $n^2\Lambda(T)$ and $E_{\textmd{\tiny{h}}}$ are in 
approximate balance, while the divergence of the conductive flux is only a 
small term (right panel).

Using the same radiative losses function, we obtain also quasi-static
intermediate temperature loops with maximum temperature $T\sim8\times10^5$~K 
(loops 7--13 in Table~\ref{tab:1}).  For loop 10, we show in the middle panels 
of Fig.~\ref{fig:TP161109} the behavior of the temperature and the pressure as 
a function of $s$ (left) and of the terms of the energy equation as a function 
of the temperature (right). The two terms competing in the energy balance are 
the divergence of the conductive flux (asterisks) and the radiative losses 
term (crosses) as in hot coronal loops. 

We explore the possibility of obtaining stable cool loops also with the 
function $\Lambda_{ANLy\alpha}$. Using the cool loop 3 of Table~\ref{tab:1} as 
a starting solution, we made a simulation by changing the radiative losses 
function from $\Lambda_{AN}$ to $\Lambda_{ANLy\alpha}$. The temperature of
loop 3 drops below $T=2\times10^4$~K (the position of the Ly-$\alpha$ peak) and 
a quasi-static, stable, cool loop of maximum temperature $1.2\times10^4$~K 
forms (loop 14 in Table~\ref{tab:1}). The temperature of this loop is 
comparable with the loop-like structures observed by VAULT but the pressure is 
ten times lower. Loop 14 is also lower and shorter with respect to loop 3. In 
order to obtain hotter loops (but still cool, with $T<10^5$~K) from loop 3 by 
changing only $E_{\textmd{\tiny{h}}}$ and using $\Lambda_{ANLy\alpha}$, we
have to raise its value from $5.5\times10^{-4}$~ergs~cm$^{-3}$~s$^{-1}$ to 
$5.8\times10^{-4}$~ergs~cm$^{-3}$~s$^{-1}$. The increase of 
$E_{\textmd{\tiny{h}}}$ should compensate the higher radiative losses (because 
of the introduction of the Ly-$\alpha$ peak) and allow the loop to keep the 
initial equilibrium without dropping to low temperatures. We obtain, indeed, a 
quasi-static cool loop with maximum temperature $T=7.4\times10^4$~K (loop 15 
in Table~\ref{tab:1}). Loop 16 of Table~\ref{tab:1} is instead obtained with 
$E_{\textmd{\tiny{h}}}=5.9\times10^{-4}$~ergs~cm$^{-3}$~s$^{-1}$ and 
$\Lambda_{ANLy\alpha}$. Using 
$E_{\textmd{\tiny{h}}}=6\times10^{-4}$~ergs~cm$^{-3}$~s$^{-1}$, the loop
reaches a maximum temperature much higher than $10^5$~K. So, loop 15 and 16 
are the two loops with temperatures higher than the temperature of the 
Ly-$\alpha$ peak that can be obtained from loop 3 changing only 
$E_{\textmd{\tiny{h}}}$ and the radiative losses function to 
$\Lambda_{ANLy\alpha}$. We are then able to ``overcome'' the Ly-$\alpha$ peak, 
whose presence was thought to prevent the existence of cool loops 
\citep{cally}. The three cool loops 14--16 have the same characteristics
already described for loops 1--6. In the bottom panels of 
Fig.~\ref{fig:TP161109}, we show the behavior of the temperature and the 
pressure as a function of $s$ (left) and of the terms of the energy equation 
as a function of the temperature (right) for loop 15. The temperature of the 
loop starts to increase at around $s=-1.1$~Mm up to $s\sim-1$~Mm, reaching the 
value of $\sim1.2\times10^4$~K and then increases rapidly till the maximum 
value of $\sim7.4\times10^4$~K. Loop 16 has the same behavior. The values of 
$L/2$ in Table~\ref{tab:1} for these two loops include the piece where the 
temperature rises slowly. The terms competing in the energy balance are the 
radiative losses term (crosses) and the background heating (solid line) as for 
loops 1--6. 

The intermediate temperature loops we have found are longer than cool
loops. Since the loop height is linked to the loop length, lower loops are 
also shorter. In the next sections, we are going to show that quasi-static 
cool loops need to be low-lying to satisfy the quasi-static equations with a  
negligible conductive flux. Intermediate temperature loops, instead, do not  
need to be as short as cool loops since their length is linked to the pressure 
and to the heating rate by RTV-like scaling laws.

\subsubsection{Loops from realistic, optically thin $\Lambda(T)$}

Using the function of radiative losses $\Lambda_{Cea}$, we obtain only
quasi-static loops with peak temperature higher than $10^5$~K (loops 18--23 in 
Table~\ref{tab:1}). Loop 17 has been obtained with the realistic radiative 
losses function $\Lambda_{Dea}$ but has the same properties of loop 18. For 
loop 17, we show in Fig.~\ref{fig:TP081010} the behavior of the temperature 
and the pressure as a function of $s$ (left panel) and of the terms of the 
energy equation as a function of the temperature (right panel). The energy 
balance terms behave as for intermediate temperature loops 7--13. 

Using $\Lambda_{Dea-H}$ and following the same method used in the case of 
$\Lambda_{ANLy\alpha}$, we obtain the cool loops 24--26 of Table~\ref{tab:1}. 
The maximum temperature of loop 24 is lower than the temperature at which 
$\Lambda_{Dea-H}$ shows the small peak (at $T\sim2\times10^4$~K) that remains 
even if the hydrogen losses have been subtracted. Loops 25 and 26 have instead 
a higher temperature. For loop 25, in the bottom panels of 
Fig.~\ref{fig:TP081010} we show the behavior of the temperature and the 
pressure as a function of $s$ (left panel) and of the terms of the energy 
equation as a function of the temperature (right panel). As for loops 15 and 
16, the temperature of loop 25 starts to increase at around $s=-5.3$~Mm 
up to $s\sim-0.8$~Mm, reaching the value of $\sim1.2\times10^4$~K and then 
increases quickly till the maximum value of $\sim4\times10^4$~K. Loop 26 
follows the same behavior. These two loops are longer with respect to loops 15 
and 16 if we consider the piece where the temperature rises slowly. We think 
that this difference is due to the different slopes of the $\Lambda(T)$ used 
in the two cases for $T<1.2\times10^4$~K. So, considering a total height of 
0.29~Mm for loop 25, only the top part with length of $\sim0.005$~Mm is hot. 
This loop is really at the limit because of its dimensions and its shape may 
depend from the details of the boundary conditions and the radiative transfer 
as we suggested previously (the shape of $\Lambda(T)$ around $T=10^4$~K). In
order to observe these kind of loops we would need at least two different 
spectral lines formed at temperatures between $10^4$ and $2\times10^4$~K in 
order to resolve the loop in its temperature extension and a very high 
resolution. As predicted by \citet{an86}, also for loops 24--26, the terms 
competing in the energy balance are only the radiative losses term (crosses) 
and the background heating (solid line). 

\subsubsection{Relations between loop parameters, and scaling laws}

In Fig.~\ref{fig:leggiscala} we show the relations between the thermodynamic 
parameters ($P$, $T_{max}$ and $L/2$) and the heating rate for the loops in
Table~\ref{tab:1} (loops 1--6 are represented by diamonds, loops 7--13 by
crosses, loops 14--16 by triangles, loops 17--23 by asterisks, and loops 
24--26 by squares). The solid lines in the lower panels of 
Fig.~\ref{fig:leggiscala} represent the RTV scaling laws for coronal loops for 
different values of $L/2$, while the dotted line (lower-right panel) 
represents the law $E_{\textmd{\tiny{h}}}=P^2$. The pressure of all the cool 
loops with $T<10^5$~K is proportional to the square root of 
$E_{\textmd{\tiny{h}}}$ and it is independent from their length and maximum
temperature. Intermediate temperature loops 7--13 and 17--23 (with different 
proportional coefficients) obey the RTV scaling laws for coronal loops. This 
is a somewhat surprising result since their temperatures are lower than 
$10^6$~K and observed intermediate temperature loops do not obey the coronal 
scaling laws \citep{brown}. The model used by \citet{rtv} to derive the 
relationships between coronal temperature, pressure, length and heating in 
coronal loops assumes a global energy balance between the heating and the 
radiative losses in static conditions. The majority of observed intermediate 
temperature loops reveal temperatures and densities that lie outside the 
static model tracks and, as different authors have shown 
\citep[e.g.,][]{dere,durrant}, the static model does not seem to accurately 
predict their physical conditions. The difference between our loops and the
observed ones is in the pressure. We obtain intermediate temperature loops
with pressures that are 1--2 orders of magnitudes lower than measured in 
observed loops with the same temperatures \citep{brown}. Loops 7--13 and 
17--23 actually have the right pressures to fall on the scaling laws lines.

\subsection{On the conditions of existence of cool 
loops}\label{sec:condexis}

The quasi-static, stable cool loops with the characteristics predicted by
\citet{an86} found are unexpected according to the considerations of 
\citet{cally}. Assuming a form of $E_{\textmd{\tiny{h}}}\propto(n/n_0)^\nu$, with 
$n_0$ the density at the loop base, the conditions of existence and stability 
of cool loops derived in the mentioned paper are not valid in the case of 
$\nu=0$ (constant heating per unit volume) combined with $a=2$. This
combination should not allow strictly static solutions. So, it is rather 
remarkable that we obtain quasi-static cool loops with the radiative losses
functions $\Lambda_{AN}$, $\Lambda_{ANLy\alpha}$, and $\Lambda_{Dea-H}$. The 
first two functions are, indeed, built with a $T^2$ dependence below $10^5$~K 
(excluding the peak for $\Lambda_{ANLy\alpha}$) and, from Fig.~\ref{fig:radlos}, 
it is clear that also $\Lambda_{Dea-H}$ follows the same behavior (excluding 
the small peak). 

$a=2$ is also empirically a better approximation to realistic $\Lambda(T)$ 
than $a=3$ used by \citet{cally}. They show that if the hydrogen losses are 
ignored, following the suggestion of \citet{mclymont} in order to obtain a 
better approximation to the true radiative losses function, then cool loops 
can exist. They assume, however, that when ignoring the hydrogen losses the 
peak in the radiative losses function is replaced by a steep slope $T^3$ for 
$T<10^5$~K. Actually, ignoring the contribution of the hydrogen in the 
function $\Lambda_{Dea}$, for example, we obtain a dependence closer to $T^2$.
Looking also at the functions $\Lambda_{Dea}$ e $\Lambda_{Cea}$ it is possible 
to say that the approximation $T^2$ below $10^5$~K is closer to the shapes of 
realistic radiative losses functions than $T^3$, and it is also often used 
\citep[see, e.g.:][]{rtv,an86}.

More surprising is the existence of cool loops even if the radiative losses
function includes a peak around $T=2\times10^4$~K. For \citet{cally}, its
presence prevents the existence of cool loops at the conditions we are 
considering ($\nu=0$) and not only. Their hydrodynamical simulations start
always from an initial uniform cool state (log$T=4.3$~K), corresponding to the 
equilibrium solution in the absence of gravity and, they say, the presence of 
the Ly-$\alpha$ plateau (they do not include the whole peak) prevents cool 
solutions, even unstable ones, with the loops reaching directly the hot state. 
All the simulations performed with a radiative losses function including the 
Ly-$\alpha$ peak ($\Lambda_{ANLy\alpha}$ and $\Lambda_{Dea-H}$), start by 
setting the quasi-static cool loop 3 as initial equilibrium state (as 
described in Sec.~\ref{subsec:description}) so that we did not have the
problems faced by \citet{cally}.

Moreover, the cool solutions we have found, not only exist, but are also 
stable \citep[contrarily to previous predictions, e.g.,][]{judge}, in the 
sense that the loop thermodynamic properties oscillate around an average 
value. These oscillations are of small amplitude and do not grow with time. 
Occasionally, however, these oscillations bring the loop to a different stable 
state, determined by the particular shape of the $\Lambda(T)$ adopted. The 
stability of these loops is therefore related to the chosen shape of 
$\Lambda(T)$ and, in particular, to the temperature value at which it reaches 
its maximum or changes slope. In the case of $\Lambda_{AN}$, if we change the 
maximum temperature value to a higher one (and increase the value of 
$E_{\textmd{\tiny{h}}}$), we can still obtain stable cool loops with a higher 
maximum temperature. As an example, we have included in Table~\ref{tab:1} 
loop 27 obtained by raising the temperature value at which $\Lambda_{AN}$ 
peaks from $T=10^{4.95}$ to $10^5$~K ($\Lambda_{AN10^5}$ in
Table~\ref{tab:1}), and using 
$E_{\textmd{\tiny{h}}}=6\times10^{-4}$~ergs~cm$^{-3}$~s$^{-1}$. The loop,
during the simulation, reaches a stable and quasi-static configuration in less 
than one hour with the maximum temperature oscillating around the value 
$\sim8.7\times10^4$~K. If we change $\Lambda_{AN10^5}$ using the original
$\Lambda_{AN}$ function, the loop reaches a new equilibrium and its maximum 
temperature increases to a value of $\sim8\times10^5$~K (loop 11 in 
Tab.~\ref{tab:1}). This happens because the random oscillations of the maximum 
temperature of the loop around its equilibrium value have sufficiently large
amplitudes to overcome, at some point, the position of the peak of the
$\Lambda(T)$ function, which for $\Lambda_{AN}$ is at log$T_{break}=4.95$~K; if 
that happens, the entire loop suddenly switches to a different equilibrium
state, corresponding to an intermediate or coronal-type loop. In 
Fig.~\ref{fig:restart}, we show the evolution of the maximum temperature of
loop 11. In the case of $\Lambda_{Dea-H}$, the maximum temperature that can be
reached by stable cool loops is related to a change in the slope ($a<2$) at 
log$T\sim4.8$~K. If we increase the value of $E_{\textmd{\tiny{h}}}$ using,
for example, loop 26 as a starting solution for a new simulation, the random 
oscillations of the maximum temperature around the equilibrium value will make 
it to overcome log$T\sim4.8$~K and the loop will find a hot equilibrium state
\citep{serio,litwin}.

The characteristics of the cool loops and, in particular, the behavior of the
three terms of the energy equation can be derived from Eqs.~\ref{eq:hydro} 
and \ref{eq:energy}. In the equations considered by \citet{an86}, the 
conductive flux is zero along the loops, so that the energy conservation 
equation (Eq.~\ref{eq:energy}) becomes $E_{\textmd{\tiny{h}}}=n^2\Lambda(T)$. 
If $a=2$, the radiative losses $n^2\Lambda(T)$ are proportional to $P^2$ 
(making use of the equation of state) and independent on the temperature, thus 
the energy balance implies that the pressure is constant. This requirement is 
obviously incompatible with pressure variations imposed by the hydrostatic 
equilibrium (Eq.~\ref{eq:hydro}). So, we need to relax the conditions of zero 
conductive flux along the loop, considering 
$|\nabla \mathbf{F}_{\textmd{\tiny{c}}}|\ll E_{\textmd{\tiny{h}}}$. In this case, we
can still have some physical solutions. Derivating Eq.~\ref{eq:energy} with 
respect to $s$ and, using the equation of state, we obtain
\begin{eqnarray}\label{eq:deriv}
\frac{\partial^2}{\partial^2 s}F_\textmd{\tiny{c}}=-2\frac{\Lambda(T)}{(2kT)^2}P\frac{\partial}{\partial s}P,
\end{eqnarray}
where we have made use of the fact that $\Lambda(T)/(2kT)^2$ is constant, if
$a=2$. In hydrostatic equilibrium we can easily estimate how small the 
required pressure gradient should be to keep the conductive flux negligible 
with respect to the other terms in the energy equation; using 
Eq.~\ref{eq:hydro} in Eq.~\ref{eq:deriv}, we have:
\begin{eqnarray}\label{eq:approx1}
\frac{\partial^2}{\partial^2 s}F_\textmd{\tiny{c}}&=&\Lambda(T)\left(\frac{P}{2kT}\right)^2\frac{2\eta(s)}{H(T)},
\end{eqnarray}
where $H(T)$ is the pressure scale height as function of temperature, defined
as $H(T)\equiv 2kT/(m_{\textmd{\tiny{H}}}g)$, and the function $\eta(s)\equiv
g_{\parallel}(s)/g(s)$ is bound in the interval [0,1]. By imposing the 
approximate equality between $E_{\textmd{\tiny{h}}}$ and radiative losses:
\begin{eqnarray}\label{eq:approx2}
\frac{\partial^2}{\partial^2 s}F_\textmd{\tiny{c}}&\simeq&
E_{\textmd{\tiny{h}}}\frac{2\eta(s)}{H(T)}.
\end{eqnarray}
An order of magnitude estimate of the loop dimension is therefore:
\begin{eqnarray}\label{eq:estimate}
\frac{\partial^2}{\partial^2 s}F_\textmd{\tiny{c}}\sim
\frac{1}{h}\left\langle\frac{\partial}{\partial s}F_\textmd{c}\right\rangle&\simeq&\frac{E_{\textmd{\tiny{h}}}}{H(\langle
  T\rangle)},
\end{eqnarray}
where the order of magnitude of the quantities is given by their averages
along the loop; factors of the order of unity have been dropped. Thus, the 
requirement that $|\nabla \mathbf{F}_\textmd{\tiny{c}}|\ll E_{\textmd{\tiny{h}}}$
implies that $h\ll H(T)$, i.e. the height of the loop must be much smaller 
than the scale height relative to the ``average'' temperature of the loop. 
Considering, for example, that the pressure scale height at $T=6\times10^4$~K 
is 3.6~Mm and $|\nabla \mathbf{F}_\textmd{\tiny{c}}|/E_{\textmd{\tiny{h}}}$ is
$\sim0.03$ for our loops, then their heights are compatible with this order of 
magnitude estimation. 

In other words, the limitation set to the solutions of Eqs. \ref{eq:hydro} and 
\ref{eq:energy} by the case $\Lambda(T)\propto T^2$, can be circumvented for 
quasi-static solution, only for low-lying (elongated) loops. Indeed, our 
solutions with temperatures below $10^5$~K are all a fraction of Mm high (the 
maximum is $\sim 0.3$~Mm, for loops 24--26 in Tab.~\ref{tab:1}).

\subsection{Calculated DEMs for cool and intermediate temperature 
loops}\label{sec:dem}

\begin{figure*}
\centering
\includegraphics[clip=true,width=13cm]{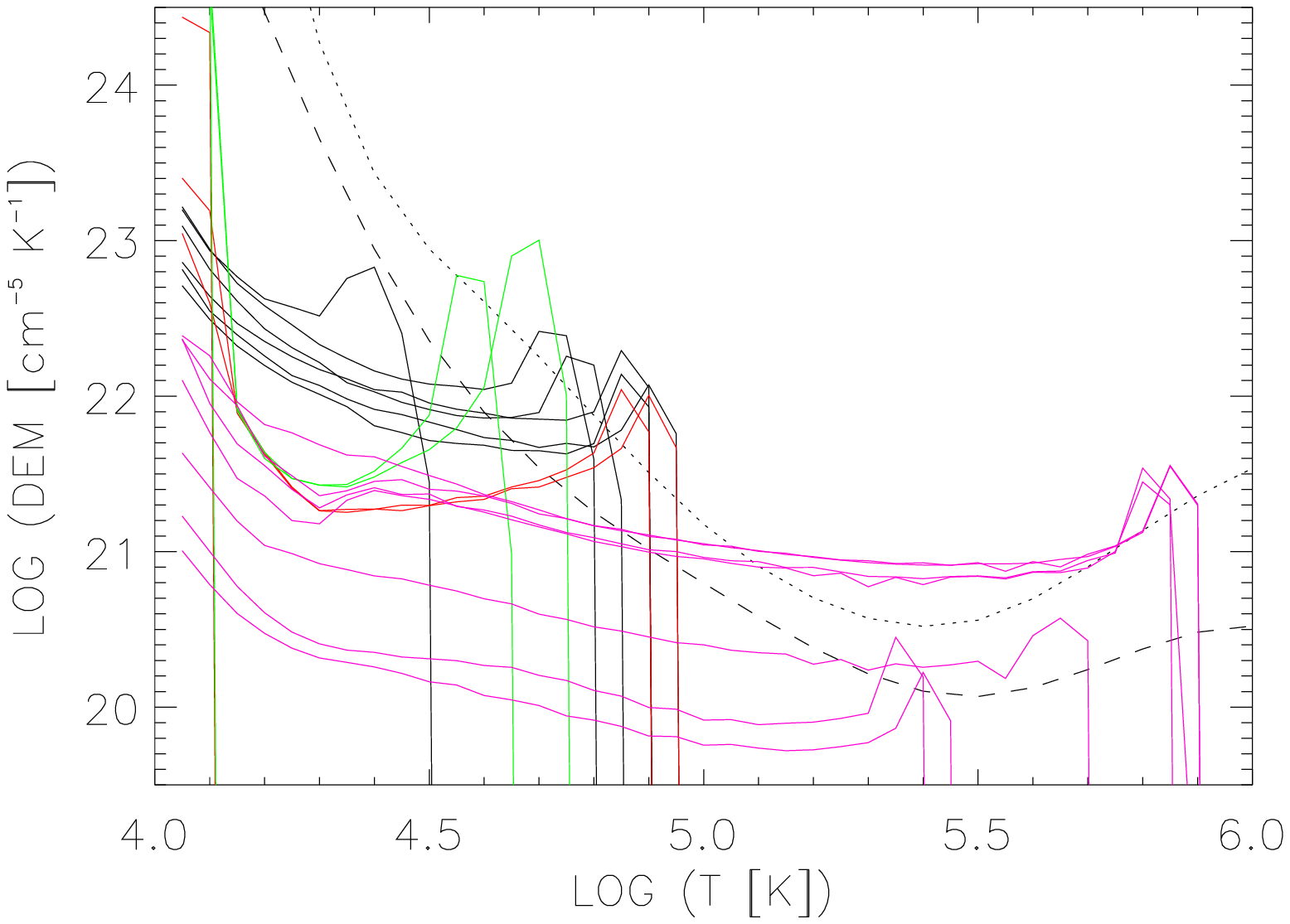}
\includegraphics[clip=true,width=13cm]{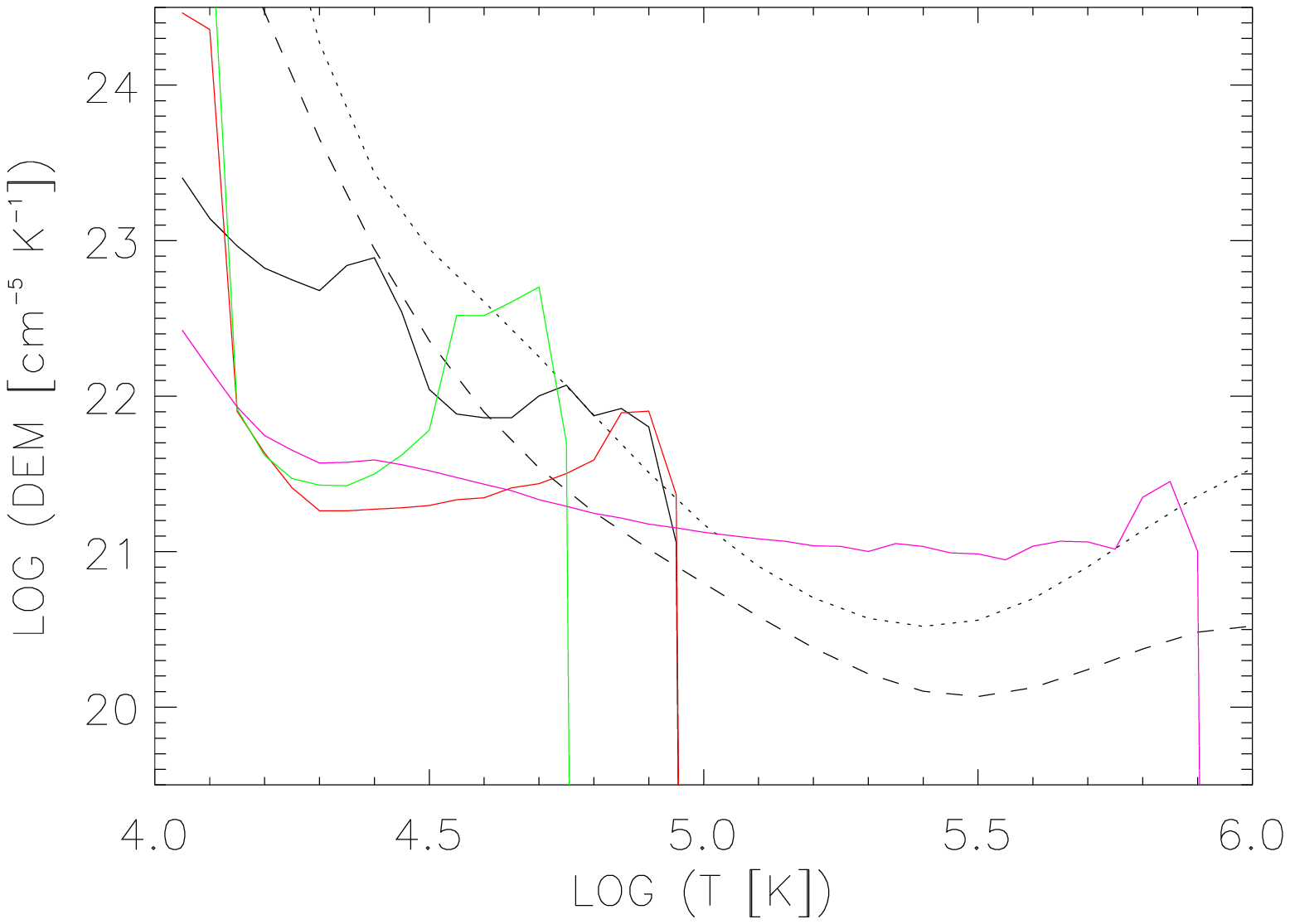}
\caption{Top: calculated DEMs for the quasi-static cool loops 1--6 (solid 
black lines), 14--16 (red), 24--26 (green) and the intermediate temperature 
loops 17--23 (magenta) of Tab.~\ref{tab:1}, compared to the DEMs of quiet sun 
(dashed) and active region (dotted) from the ``CHIANTI'' atomic data base 
\citep{chianti}. Bottom: total DEMs for each group of loops shown in the top 
panel.}
\label{fig:dem}
\end{figure*}
\begin{figure*}
\centering
\includegraphics[clip=true,width=13cm]{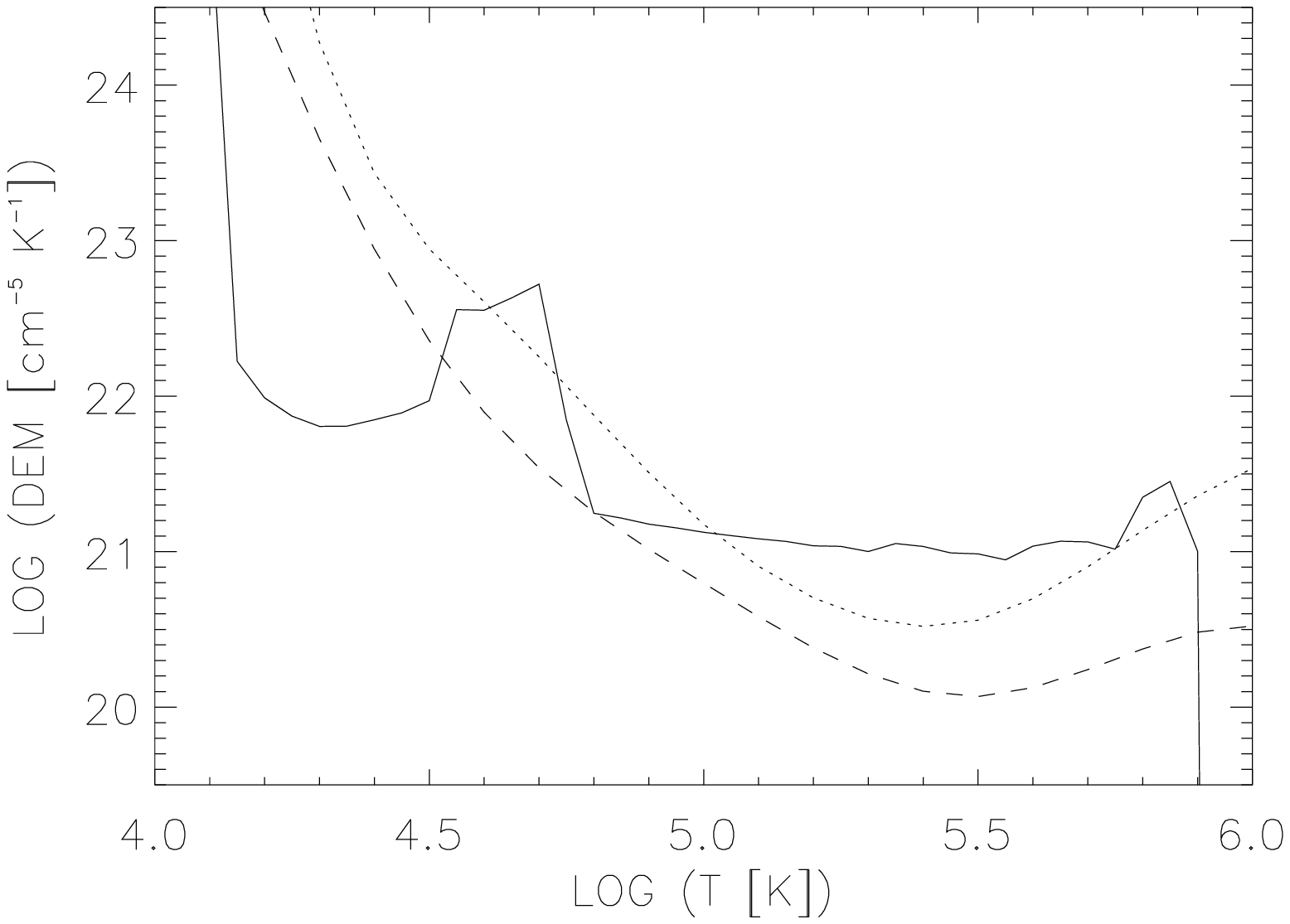}
\caption{Total DEM resulting from the combination of the DEMs of the loops 
17--23 and 24--26 (solid line), compared to the DEMs of quiet sun 
(dashed) and active region (dotted) from the ``CHIANTI'' atomic data base 
\citep{chianti}.}
\label{fig:dem3}
\end{figure*}
The theoretical DEMs for the quasi-static loops we have found are computed 
according to \citet{dem}, with a temperature bin of 0.05~dex on a $\log T$ 
scale, and considering a filling factor of $100\%$:
\begin{eqnarray}
DEM=n^2\frac{dh}{dT}.
\end{eqnarray}

The upper panel of Fig.~\ref{fig:dem} shows the calculated DEMs versus 
temperature of the quasi-static cool loops 1--6 (solid black lines), 14--16 
(red), 24--26 (green), and the quasi-static intermediate temperature loops
17--23 (magenta) of Table~\ref{tab:1}. In this figure and in the next ones,
the calculated DEMs are compared with the observed DEMs of quiet Sun and
active region (dashed and dotted lines, respectively), derived using the
\citet{vernazza} average quiet Sun and active region intensities, and produced
as part of the Arcetri/Cambridge/NRL ``CHIANTI'' atomic data base
collaboration \citep{chianti}. 

In the lower panel of Fig.~\ref{fig:dem} we plot the total theoretical DEMs, 
for each group of loops obtained with a different $\Lambda(T)$ (distinguished 
by the different colors). Assuming that the loops are equiprobable (uniformly 
distributed in log$T$) and with the same cross-section, we divide the 
temperature range in bins of amplitude $0.2$~dex on a $\log T$ scale, and 
consider for each bin a representative loop, i.e. a loop whose maximum 
temperature belongs to that bin (our loops are almost isothermal). The total
DEMs are obtained by summing the DEMs of these representative loops. When more
loops have their maximum temperature falling in the same bin, we average their 
DEMs. From Fig.~\ref{fig:dem}, we see that the largest contribution to the 
total DEM for the cool loops is given essentially by the peaks at maximum 
temperature and this contribution is determined not by the form of the single 
loop DEMs, but by the distribution of the emission measures from loop to loop 
as predicted by \citet{an86}. This is a different behavior with respect to the 
hot and the intermediate temperature loops. The DEM of a group of coronal 
loops is dominated at all temperatures by the hottest loops, having the 
largest emission measure. Hence, the DEM obtained by summing all the DEMs will 
closely resemble the DEM of the hottest loop. For a more accurate discussion 
on the shape of the resulting DEM from a group of cool and hot loops we refer 
to the calculations of \citet{an86}. They made assumptions on different 
variables (for example, the magnetic field) and loop distributions that we do 
not include in our work.

Using $\Lambda_{AN}$, we can easily obtain loops with maximum temperatures 
covering the whole temperature range, from log$T\sim4.1$~K up to the position 
of its peak (log$T=4.95$~K). Adding more DEMs of loops with different maximum 
temperatures, the resulting DEM (black solid line) can be brought to follow 
the shape of the observed ones. However, using $\Lambda_{ANLy\alpha}$ or 
$\Lambda_{Dea-H}$ the presence of the Ly-$\alpha$ peak (or a peak in general) 
at log$T\sim4.2$~K produces a relative minimum in all the DEMs, that remains 
in the total DEM (lower panel of Fig.~\ref{fig:dem}, green or red lines). We 
note that \citet{dem1} derive quiet Sun DEMs that exhibit a shape around 
log$T\sim4.2$~K that could recall the minimum that we find in our results. 

The DEMs of the intermediate temperature loops 20-23 show a minimum at a 
different temperature  with respect to the observed DEMs. We are unable to 
reproduce the whole observed DEMs considering a unique loop of this kind, in
analogy with the relationship between the DEM of a hotter loop and the
observed DEM of the corona above $10^6$~K. From Fig.~\ref{fig:dem} (lower
panel), it seems more likely that a combination of the cool and intermediate
temperature loops would assume the right shape to reproduce the observed 
emission of the lower transition region at the critical turn-up temperature 
point ($T\sim2\times10^5$~K) and below $T=10^5$~K. In Fig.~\ref{fig:dem3} we 
show indeed the total DEM (solid line) resulting from the combination of the 
DEMs of the cool loops 24--26 plus the intermediate temperature loops 
17--23. We choose to sum the DEMs of these particular cool and intermediate 
temperature loops because they have been obtained by using realistic radiative 
losses functions. Even though the $\Lambda(T)$ adopted are different, they can 
still be combined together to obtain a single DEM. We are legitimated to so 
because, while for the existence of the cool loops the shape of $\Lambda(T)$ 
under $10^5$~K is very important, we have proved instead that the exact form 
of the chromospheric radiative losses has little effect on the coronal 
properties of the loops with temperature higher than $10^5$~K. For loop 17, 
for example, changing the radiative losses function from $\Lambda_{Dea}$ to 
$\Lambda_{Dea-H}$ does not bring any change in the loop stability and 
thermodynamic parameters. This result extends what it is already known for hot 
coronal loops ($T>10^6$~K) and cooling coronal loops 
\citep[e.g.,][]{reale,brown93}.  

There is a minimum in the total DEM around log$T=4.8$~K that is due to the
lack of cool loops with that maximum temperature and $\Lambda_{Dea-H}$. This
minimum almost corresponds to the maximum of the function $\Lambda_{Dea-H}$ or
better to the point where its slope starts to change and we have $a<2$ (we have
already considered the consequences in Sec.~\ref{sec:condexis}). So, the lack 
of cool loops with maximum temperature around $\log T=4.8$~K it is not due to 
an incomplete exploration of the parameter space but to the negative slope of 
$\Lambda_{Dea-H}$ that prevents their formation. We could use the DEMs of the 
other cool loops of Table~\ref{tab:1} with maximum temperature around 
$6.5\times10^4$~K or higher but they have been obtained from idealized 
$\Lambda(T)$ and would not be suitable for a good comparison with the DEMs of 
the intermediate temperature loops 17--23. However, the shape of the averaged 
DEM of the loops 17-23, with a flat minimum and a tail extended towards low 
temperatures, helps filling this gap, improving the agreement with the 
observed DEM. Since we considered a filling factor of $100\%$ the total DEM 
has its highest value. With a lower filling factor the height of the DEM would 
be lower.

\section{Conclusions}\label{sec:concl}

We have studied the conditions of existence and stability of cool loops with 
$T\lesssim10^5$~K through hydrodynamic simulations, finding that it is 
possible to obtain quasi-static (velocities lower than $1$~km/s) cool loops, 
as predicted by \citet{an86}, stable over hours or more. These loops have been 
obtained by using different dependences of the radiative losses function on 
the temperature with respect to the work of \citet{cally}. We obtained stable 
quasi-static loops even in conditions judged prohibitive by the same authors 
on the basis of an analysis of strictly static and stationary loop equations. 
We examine and discuss the quasi-static solutions we have found, and show that 
their existence is due indeed to the small departures from static conditions, 
i.e.\ to the presence of a small but non-zero conductive flux and velocities, 
as well as to rather stringent constraints on the pressure gradients. In fact, 
for low-temperature loops, the requirement of nearly constant pressure implies 
that these loops can exist only if they are limited to small heights above the 
chromosphere (a fraction of Mm). We also show that the presence of the peak 
due to the Ly-$\alpha$ losses in the radiative losses function does not 
preclude the existence of cool loops. Moreover, we analyze the contributions 
of cool loops to the TR DEM, showing that the emission of these kind of loops 
can account for the observed DEM at $T<10^5$~K, if they were uniformly 
distributed. 

The cool loops found cannot be related with any of the observations present in
the literature because of their dimensions and especially of their small
heights. We find cool loops with lengths of 5--10~Mm, but with very low 
heights (in the range $10^{-2}-10^{-1}$~Mm). Loops so low in the solar
atmosphere (effectively embedded in the chromosphere) cannot be visible in the 
Ly-$\alpha$ line; moreover, if compared with the observations of 
\citet{patsourakos}, our loops typically have smaller pressures with respect
to their estimates.

Indeed, the shape of the cool loops we have found could give information on 
the orientation of the magnetic field in the transition region, and in 
particular on its inclination. Since, as already pointed out, these loops are 
very shallow, the magnetic flux tube should emerge at transition region 
temperatures with a quite small angle with respect to the Sun's surface and 
observations should reveal a predominance of horizontal magnetic field 
direction at their height levels, assuming that these loops are almost 
everywhere.
  
We have also obtained quasi-static loops with maximum temperature in the range
$2\times10^5-10^6$~K, using a realistic radiative losses function. These loops
are smaller with respect to coronal loops but have different characteristics 
compared to the static cool loops proposed by \citet{an86} and others. 
These loops in principle could be observed with current telescopes, but in 
order to resolve them in all their temperature extension, we would need 
multi-temperature observations, i.e. different UV lines formed at temperatures 
between $10^5-10^6$~K with resolution of at least 1''. Loops 17 and 18 have 
the maximum temperature around $2.5\times10^5$~K that is the upper limit of 
the region in which \citet{feldman1} located the ``UFS (unresolved fine 
structures)''. We find that these intermediate temperature loops follow the 
scaling laws for coronal loops contrary to results of previous works based on 
the observational data \citep[e.g.,][]{brown}. The loops obtained have indeed 
small pressures that make their parameters obey the RTV scaling laws, but 
these pressures are 1--2 orders of magnitudes lower than the ones estimated 
from the observations \citep{brown}. Observed intermediate temperature loops 
are usually associated with flows while we are studying possible solutions to 
the hydrodynamic equations in quasi-static conditions without considering 
their role. The small values of the pressure we have found for these loops are 
the only possible in order to find solutions to the hydrodynamic equations in 
quasi-static conditions. Due to their small pressures, these loops obey to the 
scaling laws for coronal loops. The presence of substantial flows could 
influence their pressures. 
 
We are not able to reproduce the DEMs derived from observations with only one 
set of parameters (a single loop). A similar problem has already been reported 
in the paper of \citet{susino} but for coronal loops. We find instead that a 
combination of cool and intermediate temperature loops, in particular 
precisely due to their computed pressures, can give a DEM with a shape not too 
far from observed. This of course does not preclude the possibility that some 
additional physical processes, not included in our simulations (like that 
proposed by \citealt{judge1} or by \citealt{klimchuk1} and 
\citealt{depontieu}, or dynamic structures like cooling, heating loops and 
spicules), must be taking place in the transition region and contribute to 
explain the discrepancies with the observed DEM structure. Due to their 
dimensions, the emission of the quasi-static loops found can be seen as a 
diffuse component of the transition region.

In our simulations, we have considered only the case of constant heating rate 
and we have already found quasi-static cool loops in conditions not allowed
from strictly static and stationary hydro-dynamical equations. More realistic 
assumptions could make obtaining stable, quasi-static cool loops even
easier. In particular, in this work we have shown how the shape of the
radiative losses function below $10^5$~K is very important for the existence
of cool loops. Because the radiative losses around log$T\sim4.2$~K are 
dominated by the \ion{H}{i} Ly-$\alpha$, we plan to explore in a follow-up
study the effect on the structure and stability of cool loops of a more
realistic treatment of hydrogen radiation losses in the lower TR. To do so, it 
will be necessary to improve the models by introducing a full treatment of
optically thick radiative losses and/or partial ionization, and therefore
solving the radiative transfer equations and rate equations of hydrogen
(non-local thermodynamic equilibrium). Our first attempt to simulate an 
optically thick radiative losses function by removing the hydrogen losses from 
a realistic radiative losses function seems to be promising. Without the 
hydrogen losses, the function $\Lambda_{Dea-H}$ becomes lower and we are able 
to obtain quasi-static cool loops with $T<10^5$~K.

Taking into account finite cross sections for these loops could influence the 
details of the radiative transfer in the Ly-$\alpha$ line (by reducing, for 
instance, the escape probability of Ly-$\alpha$ photons). Predicting the 
effect of loop area expansion above the chromosphere is not so easy, although 
it is difficult to envision substantial expansion factors in such small loops.

Finally, the maximum temperature reachable by cool loops is limited only by 
the second maximum (or a change of slope) present in the $\Lambda(T)$ since we 
have shown that the peak due to the Ly-$\alpha$ losses is not a limit. The 
different radiative losses functions used have this maximum (or the slope 
change) at different temperatures. So, we underline the importance of knowing 
accurately the $\Lambda(T)$ also at temperature around $10^5~K$ for the 
structure of the cool loops and of what we termed ``intermediate loops'', in 
analogy with the importance of the position and strength of the Ly-$\alpha$ 
peak for cool loops. 

\begin{acknowledgements}
This work was supported by the ASI/INAF contracts I/05/07/0 for the program 
``Studi Esplorazione Sistema Solare'', and I/023/09/0 for the program ``Attivit\`a scientifica per l'analisi dati Sole e plasma - Fase E2/F''.
\end{acknowledgements}

\bibliographystyle{aa}

\end{document}